\documentclass[11pt,a4paper]{article}
\usepackage{jheppub}
\pdfoutput=1

\usepackage[utf8]{inputenc}
\usepackage{amssymb}
\usepackage{amsmath}
\usepackage{amsfonts}
\usepackage{graphicx}
\usepackage{xcolor}
\usepackage{xspace}
\usepackage[normalem]{ulem} %\usepackage[normalem]{ulem} if want italics
\usepackage{lscape}
\usepackage{ wasysym }
\usepackage{float}
\usepackage{mathrsfs}
\usepackage{slashed}
\usepackage{multirow,rotating}

\usepackage{hyperref}
\hypersetup{
  colorlinks=true,
  allcolors=darkpurple,
  pdfborder={0 0 0},
  linktocpage=true%false
}

\definecolor{darkpurple}{rgb}{0.5,0,0.5}
\definecolor{cambridgeblue}{rgb}{0.64, 0.76, 0.68}
\definecolor{darkraspberry}{rgb}{0.53, 0.15, 0.34}

\def\gsim{\raise0.3ex\hbox{$\;>$\kern-0.75em\raise-1.1ex\hbox{$\sim\;$}}}
\def\lsim{\raise0.3ex\hbox{$\;<$\kern-0.75em\raise-1.1ex\hbox{$\sim\;$}}}

\newcommand{\ba}[1]{\begin{eqnarray} \label{(#1)}}
\newcommand{\ea}{\end{eqnarray}}

\def\gsim{\raise0.3ex\hbox{$\;>$\kern-0.75em\raise-1.1ex\hbox{$\sim\;$}}}
\def\lsim{\raise0.3ex\hbox{$\;<$\kern-0.75em\raise-1.1ex\hbox{$\sim\;$}}}

\makeatletter
\g@addto@macro\bfseries{\boldmath}
\newcommand\Label[1]{&\refstepcounter{equation}(\mathrm{\theequation})\ltx@label{#1}&}
\makeatother

\graphicspath{{figs/}}

\allowdisplaybreaks

\preprint{IFIC/23-04}	
\title{Reinterpretation of searches for long-lived particles from meson decays}

\author[a]{Rebeca Beltr\'an,}
\emailAdd{rebeca.beltran@ific.uv.es}
\affiliation[a]{{\it AHEP Group, Instituto de F\'{\i}sica Corpuscular --	CSIC/Universitat de Val{\`e}ncia, Apartado 22085,
	E--46071 Val{\`e}ncia, Spain}}

\author[b,c]{Giovanna Cottin,}
\emailAdd{giovanna.cottin@uai.cl}
\affiliation[b]{Departamento de Ciencias, Facultad de Artes Liberales, 
	Universidad Adolfo Ib\'a\~{n}ez,
	Diagonal Las Torres 2640, Santiago, Chile}
\affiliation[c]{Millennium Institute for Subatomic Physics at the High Energy Frontier (SAPHIR), Fernández Concha 700, Santiago, Chile}

\author[a]{Martin Hirsch,}
\emailAdd{mahirsch@ific.uv.es}

\author[d]{Arsenii Titov,}
\emailAdd{arsenii.titov@df.unipi.it}
\affiliation[d]{Dipartimento di Fisica ``Enrico Fermi'', 
Università di Pisa and INFN, Sezione di Pisa,
Largo Bruno Pontecorvo 3, I--56127 Pisa, Italy}

\author[e,f]{Zeren Simon Wang}
\emailAdd{wzs@mx.nthu.edu.tw}
\affiliation[e]{Department of Physics, National Tsing Hua University, Hsinchu 300, Taiwan}
\affiliation[f]{Center for Theory and Computation, National Tsing Hua University, Hsinchu 300, Taiwan}

\abstract{Many models beyond the Standard Model predict light and feebly interacting particles that are often long-lived.
These long-lived particles (LLPs) in many cases can be produced from meson decays.
In this work, we propose a simple and quick reinterpretation method for models predicting LLPs produced from meson decays.
With the method, we are not required to run Monte-Carlo simulation, implement detector geometries and efficiencies, or apply experimental cuts in an event analysis, as typically done in recasting and reinterpretation works.
The main ingredients our method requires are only the theoretical input, allowing for computation of the production and decay rates of the LLPs.
There are two conditions for the method to work: firstly, the LLPs in the models considered should be produced from a set of mesons with similar mass and lifetime (or the same meson) and second, the LLPs should, in general, have a lab-frame decay length much larger than the distance between the interaction point and the detector.
As an example, we use this method to reinterpret exclusion bounds on heavy neutral leptons (HNLs) in the minimal ``3+1'' scenario, into those for HNLs in the general effective-field-theory framework as well as for axion-like particles.
We are able to reproduce existing results, and obtain new bounds via reinterpretation of past experimental results, in particular, from CHARM and Belle.
}

\begin{document}
\maketitle

%%%%%%%%%%%%%%%%%%%%%%%%%%%%%%%%%%%%%%%%%%%%%%%%%%%%%%%%%%%%%%%%%%%%%%
%\tableofcontents
%
%%%%%%%%%%%%%%%%%%%%%%%%%%%%%%%%%%%%%%%%%%%%%%%%%%%%%%%%%%%%%%%%%%%%%%

%\newpage

% !TEX root = ../reinterpretation.tex

\section{Introduction}\label{sec:intro}

With a lack of concrete discovery of heavy new fields inspired mainly by supersymmetry~\cite{Nilles:1983ge,Martin:1997ns} at the LHC, increasingly more attention has been shifted in recent years towards light and feebly interacting new physics (NP).
Such exotic form of NP often involves long-lived particles (LLPs)\footnote{See Refs.~\cite{Alimena:2019zri,Lee:2018pag,Curtin:2018mvb,Knapen:2022afb} for some recent summary reviews.} that perhaps have so far escaped LHC searches.
LLPs are predicted in a wide range of models beyond the Standard Model (BSM), and usually motivated for solving issues in the Standard Model (SM) of particle physics, such as the non-vanishing neutrino mass and the nature of dark matter (DM).
For instance, a class of the so-called ``portal-physics'' models predict heavy neutral leptons (HNLs), dark photons, dark Higgs, and axion-like particles (ALPs), that can be either promptly decaying or long-lived, and may connect our visible SM sector to a hidden sector of the DM.

LLPs can be produced via different channels at high-energy colliders, 
$B$-factories, beam-dump experiments, and so on.
For instance, they can be directly produced via beam collisions, \textit{e.g.}~electroweakino production at proton-proton colliders~\cite{CMS:2020atg,ATLAS:2022rme}.
LLPs can also originate from (rare) decays of mesons, electroweak gauge bosons, Higgs bosons, or perhaps heavy new particles.
Testing all the different channels can usually allow for probing complementary parts of the model parameter space.
In this study, we will focus on LLPs produced from mesons.

In fact, there exist a wide range of models predicting LLPs that can be produced from meson decays, such as HNLs, ALPs, and the lightest neutralino in R-parity-violating supersymmetry~\cite{Dedes:2001zia,Dreiner:2002xg,Dreiner:2009er,deVries:2015mfw}.
When an experiment publishes its results for a search, usually a certain model, say the HNLs in a minimal ``3+1'' scenario, is chosen to be interpreted for the experimental results.
If the experimental search can, in principle, also constrain another model, traditionally a proper reinterpretation would require knowledge of the detailed signal-event chain topologies, of the main beam collisions,  of the detector geometries and efficiencies, and of the cuts taken by the search. For instance, the recasting tool CheckMATE~\cite{Drees:2013wra,Kim:2015wza,Dercks:2016npn} can call programs such as MadGraph5~\cite{Alwall:2014hca} or Pythia8~\cite{Sjostrand:2014zea} to perform Monte-Carlo (MC) simulation including hadronization and showering, and Delphes3~\cite{deFavereau:2013fsa} for fast detector simulation, followed by applying closely the experimental event-selection cuts, to determine if a parameter point of a model is excluded or allowed by the search.
Or the tool SModelS~\cite{Kraml:2013mwa,Khosa:2020zar} studies simplified model spectrum topologies and confronts them with the related experimental bounds, with no necessity to perform MC simulation.

For LLP searches, there are additional challenges for reinterpretation, as there are no standard definitions for LLP objects nor enough experimental information publicly available for LLP reconstruction in most cases~\cite{LHCReinterpretationForum:2020xtr,Alimena:2019zri,Proceedings:2018het}.
Current efforts for addressing LLP reconstruction within standard tools include CheckMATE~\cite{Desai:2021jsa}, MadAnalysis~\cite{Araz:2021akd}, and SModelS~\cite{Alguero:2021dig}.
Other recasting LLP simulation efforts include public repositories with specific long-lived particle searches as displaced vertices or disappearing charged tracks~\cite{LLPrecastingRepo}, as well as a dedicated new Delphes module to reconstruct displaced showers~\cite{delphes_pr}. 

Here, we wonder, whether it is possible to conduct reinterpretation of LLPs from \textit{meson decays} without running simulation, which is time-consuming, and also with no need to check and compare the details of the event topologies, nor the knowledge of the relevant collision and detector information; can we accomplish the job with only a computation of the LLPs' production and decay rates?
We find this is achievable under certain conditions, \textit{i.e.}~if the LLPs in different models have similar kinematics, \textit{e.g.}~all are produced from a set of mesons with similar mass and lifetime or the same meson (at the identical experiment), and have a lab-frame decay length much larger than the distance between the interaction point (IP) and the detector of the experiment.

It is impractical to consider all the possible models in one paper, and therefore we choose to restrict ourselves to the HNLs and ALPs only, for illustrative purposes.
We will now briefly discuss the models we consider here.
Firstly, HNLs are perhaps the most hunted candidate of LLPs, which are highly motivated for various reasons.
They are fermionic SM singlets that mix with one, two, or all three species of the SM active neutrinos.
They can explain the non-vanishing active neutrino masses in an elegant way by the so-called seesaw mechanism and its variations~\cite{Minkowski:1977sc,Yanagida:1979as,Gell-Mann:1979vob,Mohapatra:1979ia,Schechter:1980gr,Wyler:1982dd,Mohapatra:1986bd,Bernabeu:1987gr,Akhmedov:1995ip,Akhmedov:1995vm,Malinsky:2005bi}.
In the minimal ``3+1'' scenario of the HNLs, there are only two types of parameters in the model, \textit{i.e.}~the HNL mass and mixing angles, controlling both production and decay of the HNLs.
Oftentimes for the purpose of performing phenomenological studies, the HNLs are assumed to mix with only one generation of the active neutrinos, and hence constraints are obtained in a plane spanned by two independent parameters, the mixing angle and the mass.
Even though this choice is unrealistic as neutrino oscillation data 
(see Refs.~\cite{Capozzi:2021fjo,Esteban:2020cvm,deSalas:2020pgw} for recent global analyses) requires at least two massive HNLs, 
it is a useful minimal choice to characterize a possible experimental signal.
At beam-dump experiments such as CHARM~\cite{CHARM:1985nku}, HNLs can arise from weak decays of various mesons that are copiously produced there.
$B$-factories including Belle~\cite{Belle:2013ytx} and BaBar~\cite{BaBar:2022cqj} have also obtained bounds for HNLs from $B$-meson decays.
Moreover, in recent years, a series of far-detector programs have been proposed (with some already approved)
to be operated in the vicinity of different IPs at the LHC, aiming mainly to detect displaced-vertex signatures of LLPs.
These include FASER~\cite{Feng:2017uoz,FASER:2018eoc}, MATHUSLA~\cite{Curtin:2018mvb,Chou:2016lxi,MATHUSLA:2020uve}, and MoEDAL-MAPP~\cite{Pinfold:2019nqj,Pinfold:2019zwp}, among others.
Since the production rates of mesons at the LHC are huge, these experiments are estimated to be highly sensitive to LLPs from meson decays.
Phenomenological studies on the exclusion limits of these far detectors on the HNLs in the minimal case can be found \textit{e.g.}~in Refs.~\cite{Helo:2018qej,Kling:2018wct,Hirsch:2020klk,Curtin:2018mvb,Ovchynnikov:2022its,Aielli:2019ivi,DeVries:2020jbs}.

Beyond the minimal scenario, HNLs also appear in other BSM models that often extend the SM with new gauge groups.
Examples include models with an extra $U(1)$ gauge group predicting a new gauge boson $Z'$~\cite{Chiang:2019ajm}, and leptoquark models predicting heavy leptoquark particles~\cite{Dorsner:2016wpm}.
In this class of models, the production and decay of the HNLs are usually mediated by two independent parameters, respectively, besides the HNL mass.
If HNLs have masses around the weak scale and other new particles are much heavier, the effects of NP at low energies can be described in terms of
the Standard Model effective field theory extended with HNLs, known as $N_R$SMEFT~\cite{delAguila:2008ir,Aparici:2009fh,Liao:2016qyd}.
In the $N_R$SMEFT, one writes down non-renormalizable operators including one or more HNLs in the Lagrangian and the operator coefficients are associated with the NP scale, usually labeled as $\Lambda$.
For GeV-scale HNLs produced in meson decays, the appropriate description is provided by the low-energy effective field theory extended with HNLs, $N_R$LEFT~\cite{Bischer:2019ttk,Chala:2020vqp,Li:2020lba,Li:2020wxi}, in which heavy SM degrees of freedom (namely, the top quark, the Higgs and $Z$ and $W^\pm$ gauge bosons) are not present.
In this paper, we will adopt this framework and will show how the bounds on the HNLs in the minimal scenario can be translated to those on the Wilson coefficients of the $N_R$LEFT.
For simplicity, for the HNLs in either the minimal scenario or effective-field-theory (EFT) scenarios, we will study their association with a charged lepton or active neutrino of the electron flavor only.

Besides the HNLs, there are other possible LLPs that may be produced from meson decays, such as ALPs.
ALPs are the pseudo-Nambu-Goldstone bosons associated with 
an approximate global symmetry that is spontaneously broken at a high scale $\Lambda$.
They are mainly motivated as DM candidates~\cite{Preskill:1982cy,Abbott:1982af,Dine:1982ah,Panci:2022wlc}, and their mass and couplings are independent and can range across orders of magnitude~\cite{Kim:2015yna,DeMartino:2017qsa,Rubakov:1997vp}. 
At low energies, the ALP interactions are described by the corresponding EFT~\cite{Georgi:1986df,Choi:1986zw}. 
This theory has gained significant attention in recent years, see \textit{e.g.}~Refs.~\cite{Brivio:2017ije,Chala:2020wvs,Bauer:2020jbp,Bauer:2021mvw,MartinCamalich:2020dfe,Calibbi:2020jvd}, in part because of the lack of observation of NP at the LHC.
The (pseudo-)Goldstone nature of the ALP manifests itself in its derivative couplings to SM fermion currents.
In particular, the flavor off-diagonal couplings to up (down) quarks can trigger the decays of $D$($B$)-mesons to a lighter meson and the ALP.
The latter can be long-lived, decaying \textit{e.g.} via its couplings to charged leptons, cf.~Ref.~\cite{Dreyer:2021aqd}.
In this work, we will show how the searches for HNLs in the minimal scenario can constrain the parameter space of the ALP EFT.

Since all these LLPs are produced from meson decays, they should have similar kinematics, as long as they come from a set of similar mesons such as $D$-mesons ($D^0$, $D^\pm$ and $D_s$) or only $B$-mesons ($B^0$, $B^\pm$ and $B_s$), or even the same meson, at the same experiment.
This is legitimate because the LLPs stemming from mesons of similar masses and lifetimes are expected to have similar boost factor and polar angle distributions and hence similar decay probabilities in the detector, given the same LLP proper lifetime and mass.\footnote{The spin of the LLP, the number of LLPs in each signal meson decay, as well as whether the meson decay is two-body or three-body, should have an impact on the distribution. However, as we will see when we discuss numerical results, this impact is small. Therefore, we do not discuss it in detail here.}
Moreover, from the experimental exclusion limit point of view, the small coupling regime (or heavy NP scale) is almost always harder to probe than the large coupling one (or light NP scale), and in the small coupling regime, the LLPs are usually very long-lived given their small masses as required by kinematic constraints (production from meson decays).
Therefore, our requirements for the simple reinterpretation as mentioned above are fulfilled most of the time, for the LLPs originating from mesons.
In this work, as a proof-of-principle example, we will show how to reinterpret bounds on the minimal HNL scenario (in the plane mixing angle vs.~mass) produced from charm and bottom meson decays, into the model-parameter planes of (i) the $N_R$LEFT and (ii) the ALP EFT.
For the $N_R$LEFT, we will consider operators that include either one or two HNLs, \textit{i.e.}~single-$N_R$ or pair-$N_R$ operators.

In the next section, we elaborate on the reinterpretation method we propose for LLPs produced from meson decays.
Sec.~\ref{sec:experiments} contains a brief introduction to both the past and future experiments we consider, as well as relevant experimental bounds for the theoretical scenarios we are studying in this work.
Then in Sec.~\ref{sec:models}, we show the results of reinterpreting the sensitivity limits on HNLs in the minimal scenario into HNLs in the EFT and into ALPs.
Finally, we discuss the advantages and limitations of the proposed reinterpretation method and provide an outlook, in Sec.~\ref{sec:discussionANDsummary}.

% !TEX root = ../reinterpretation.tex
\section{Reinterpretation method}\label{sec:method}

To explain the reinterpretation procedure, we start with a general discussion on searches for the HNLs in the minimal scenario.
First, once an experimental search for the minimal HNL scenario is finished, say, without a discovery, the $95\%$ (or sometimes $90\%$) confidence level (C.L.) exclusion limits corresponding to a certain number of signal events, $N_S$, are determined according to the numbers of observed events and expected background events.
These bounds can be converted to model parameters with the following formula,
\begin{eqnarray}\label{eqn:NS_minimal}
N_S = N_N \cdot \epsilon \cdot \text{BR}(N\to\text{vis.})\,,
\end{eqnarray}
where $N_N$ is the total number of HNLs produced from certain meson decays, $\epsilon$ is the product of the detector acceptance and efficiencies, and $\text{BR}(N\to\text{vis.})$ is the decay branching ratio of the HNLs into certain final states visible/detectable/observable in the detector chamber.
$N_N$ and BR$(N\to\text{vis.})$ can just be computed from the model parameters, mixing angle squared $|V_{eN}|^2$ and mass $m_N$, and $\epsilon$ should be determined by a proper simulation together with the computation of the proper lifetime of the HNL, $c\tau_N$, from $|V_{eN}|^2$ and $m_N$.
Thus, the experimental search can obtain bounds in the plane $|V_{eN}|^2$ vs.~$m_N$.
Therefore, from the published results of the experimental search, we obtain a three-dimensional (3D) array of data in $(m_N, |V_{eN}|^2, N_S)$, with $N_S$ fixed at a certain single value (\textit{e.g.}~$N_S=3$ for $95\%$ C.L.~exclusion limits with zero observed event) and $m_N$ and $|V_{eN}|^2$ varying.
Alternatively, if we could perform the simulation including the experimental event-selection cuts by ourselves, we could also obtain another set of 3D data for the same set of variables, but with $|V_{eN}|^2$ fixed at a certain value, and $m_N$ and $N_S$ varied.

Now, assuming the HNLs are in the large decay length limit, such that its boosted decay length, $\lambda_{\text{decay}}=\beta\gamma c\tau=\beta\gamma c \hbar/\Gamma_{\text{tot.}}$, with $\Gamma_{\text{tot.}}$ being the HNL total decay width, is much larger than the distance from the detector to the IP, $\epsilon$ would be simply linearly proportional to the total decay width of the HNL, $\Gamma_{\text{tot.}}$ in general.
This can be understood as follows.
$\epsilon$ is proportional to the detector acceptance to the HNL, $P[\text{decay}]$, which can be essentially expressed with the following formula:
\begin{eqnarray}
 P[\text{decay}]&\sim&\text{exp}(-L/\lambda_{\text{decay}}) \cdot ( 1  - \text{exp}(-\Delta L/\lambda_{\text{decay}})) \nonumber\\
& =& \text{exp}(-L/\lambda_{\text{decay}}) - \text{exp}(-(L+\Delta L)/\lambda_{\text{decay}}),\nonumber\\
&\approx& \Delta L/\lambda_{\text{decay}} 
= \Delta L \cdot \Gamma_{\text{tot.}}/(\beta\gamma c \hbar)\,,\label{eq:decayprob}
\end{eqnarray}
where $L$ is the distance from the IP to the detector, $\Delta L$ is the length inside the detector that the HNL traverses if it does not decay inside the detector, and the second-last equality holds for $\lambda_{\text{decay}}\gg L$. 
One sees that in the large decay length limit, the exponential decay distribution can be approximated as linearly proportional to the HNL total decay width.
As a result, we can derive from Eq.~\eqref{eqn:NS_minimal} the following relation:
\begin{eqnarray}\label{eqn:NS_minimal_simplifed}
N_S \propto   N_N \cdot \Gamma_{\text{tot.}} \cdot \text{BR}(N\to\text{vis.}) = N_N\cdot \Gamma_{\text{vis.}}\,,
\end{eqnarray}
where the last equality stands because $\text{BR}(N\to\text{vis.})= \Gamma_{\text{vis.}} / \Gamma_{\text{tot.}}$ where $\Gamma_{\text{vis.}}$ denotes the sum of partial widths of the HNLs into visible final states.
Now considering the HNLs in the $N_R$LEFT, we can write a similar relation,
\begin{eqnarray}\label{eqn:NS_EFT_simplified}
N'_S \propto N'_N\cdot \Gamma'_{\text{vis.}}\,,
\end{eqnarray}
where the primed variables are the EFT-counterparts of those in Eq.~\eqref{eqn:NS_minimal_simplifed}.
We assume for now that the production and decay Wilson coefficients are different in the $N_R$LEFT.
Note that the same relation can also be written for other types of LLPs than the HNLs, such as the ALPs, but we will stick to the HNLs in the EFT in the rest of this section, for convenience of discussion.

Here, we should comment that as long as we work in the large decay length limit, and the HNLs in both cases are produced from the same type of mesons ($D$-mesons only or $B$-mesons only, for instance), the kinematics of the HNLs in the two cases are sufficiently similar so that we can ignore the differences for our purpose and assume that the proportionality is shared between Eq.~\eqref{eqn:NS_minimal_simplifed} and Eq.~\eqref{eqn:NS_EFT_simplified}.
This allows us to take the ratio of Eq.~\eqref{eqn:NS_minimal_simplifed} to Eq.~\eqref{eqn:NS_EFT_simplified} and to reach the following relation:
\begin{eqnarray}\label{eqn:NSrelation}
\frac{N_S}{N'_S} \approx  \frac{N_N}{N'_N}\frac{\Gamma_{\text{vis.}}}{\Gamma'_{\text{vis.}}}\,,
\end{eqnarray}
or equivalently,
\begin{eqnarray}\label{eqn:ctaurelation}
\Gamma'_{\text{vis.}} \approx \Gamma_{\text{vis.}} \cdot \frac{N_N}{N'_N} \frac{N'_S}{N_S}\,.
\end{eqnarray}
We further note that in a commonly seen case where $N_S=N'_S$, Eq.~\eqref{eqn:ctaurelation} can be simplified to be:
\begin{eqnarray}\label{eqn:ctaurelation_simplified}
\Gamma'_{\text{vis.}} \approx \Gamma_{\text{vis.}} \cdot \frac{N_N}{N'_N}\,.
\end{eqnarray}

As discussed above, we already have a list of 3D data in $(m_N, |V_{eN}|^2, N_S)$.
For each $m_N$, the corresponding value of $|V_{eN}|^2$ allows to compute $N_N$ and $\Gamma_{\text{vis.}}$.
Further, $N_S$ is known, and $N'_S$ is usually either equal to $N_S$ or can be determined from the experimental search depending on the final states of the LLP decays.
As a result, for the $N_{R}$LEFT case, once we fix the production Wilson coefficient determining $N'_N$, we can derive $\Gamma'_{\text{vis.}}$ with the usage of Eq.~\eqref{eqn:ctaurelation} or Eq.~\eqref{eqn:ctaurelation_simplified}.
It is then a straightforward exercise to convert these bounds on $\Gamma'_{\text{vis.}}$ into the corresponding bounds on the decay Wilson coefficient in the $N_{R}$LEFT.

We stress again this approximation holds only if in both the minimal scenario and the $N_R$LEFT scenario:
\begin{enumerate}
	\item the HNLs are produced from the same type of mesons, ensuring the same kinematics, and
	\item the HNLs are long-lived relative to the detector distance from the IP, ensuring working in the linear regime for the exponential decay distributions.
\end{enumerate}

In the procedure above, it has been assumed that the production and decay couplings of the LLP in the new model (the ALP and HNL in the EFTs as in this work) are unrelated.
However, it is also often the case that they are related.
Labeling these two Wilson coefficients as $P_{\text{prod}}$ and $P_{\text{decay}}$, we discuss briefly here, without loss of generality, the case that they are equal: $P_{\text{prod}}=P_{\text{decay}}$.
We first fix a value for $P_{\text{prod}}$ and derive a limit on $P_{\text{decay}}$ following the algorithm given above.
In the general case, the derived value of $P_{\text{decay}}$ is unequal to $P_{\text{prod}}$.
We notice the fact that if we chose a value of $P_{\text{prod}}$ larger (smaller) by a random positive number $x$, we would obtain a bound on $P_{\text{decay}}$ stronger (weaker) also by $x$, assuming the production and decay rates of the LLP are proportional to $P_{\text{prod}}^2$ and $P_{\text{decay}}^2$, respectively.
This means if we multiply $P_{\text{prod}}$ by $\sqrt{P_{\text{decay}}/P_{\text{prod}}}$ to reach the geometric mean of $P_{\text{prod}}$ and $P_{\text{decay}}$, this new production coupling would lead to a new bound on $P_\text{decay}$ of equal value.
We emphasize again that all the manipulation here works only if the boosted decay lengths of the LLPs in the relevant parts of the parameter space are sufficiently large.

% !TEX root = ../reinterpretation.tex
\section{Experiments}\label{sec:experiments}

In this section, we provide details of existing searches to be recast
with our reinterpretation method, as well as comment on existing upper limits on the branching ratios of (semi-) invisible decays of $D$- and $B$-mesons.
The latter will provide strong constraints on the couplings in our selected benchmark scenarios. 
Additionally, we briefly describe the future LLP far detectors, whose sensitivities to the minimal HNL scenario will be reinterpreted into the $N_{R}$LEFT and the ALP EFT in Sec.~\ref{sec:models}.

% !TEX root = ../reinterpretation.tex
\subsection{Past experiments}\label{subsec:past_exp}

Several past experiments can place constraints on HNLs and ALPs (see \textit{e.g.}~Refs.~\cite{Boiarska:2021yho,Barouki:2022bkt}). A summary of bounds in the minimal HNL scenario can be found \textit{e.g.}~in Ref.~\cite{hnl_limits}.
In this section, we concentrate on three past experiments, which can provide stringent bounds on LLP masses between roughly 0.1 and 5 GeV.  We discuss each particular case below.

\begin{itemize}

\item \textbf{CHARM:} HNLs can be produced from $D$-mesons in semi-leptonic decays (from charged and neutral $D$'s) and via  leptonic decays of charged $D$'s, see \textit{e.g.}~Ref.~\cite{Bondarenko:2018ptm}.
For the minimal HNL scenario with HNL mixing in the electron sector, and for HNL masses between the kaon and the $D$ mass, the strongest constraint comes from the CHARM beam-dump experiment~\cite{CHARM:1983ayi,CHARM:1985nku}, where a prompt neutrino beam was produced at CERN SPS by dumping 400 GeV protons on a thick copper target.
Searches for decays of HNLs were performed within a detector decay region of 35~m length and 9~m$^2$ surface area.
The detector was located 480 m away from the copper target. 
The values of active-heavy mixing $V_{eN}$ as small as $|V_{eN}|^2\sim 10^{-7}$ for $m_{N}\approx 1$~GeV were probed.

In the experimental analysis~\cite{CHARM:1985nku}, HNLs were searched for in the leptonic decays of $D^{\pm}$ and semi-leptonic decays of both $D^{\pm}$ and $D^{0}$. We consider all two and three body decays of both $D^{0}$ and $D^{\pm}$,\footnote{We note that Ref.~\cite{CHARM:1985nku} assumed a ratio of $\sigma(D^{\pm})=\sigma(D^{0})/2$ in the estimates of the HNL production. We follow this choice when computing $N_N$ and $N'_N$ for both the EFT HNL and ALP scenarios. In addition, Ref.~\cite{Boiarska:2021yho} claims that HNL production from $D_s$ decays was not included in the CHARM search~\cite{CHARM:1985nku}.
We therefore neglect the $D_s$ contributions, which would be negligible anyway for the relatively small fragmentation factor of $c\to D_s$.}
where $N$ decays further via mixing in the electron sector 
as i) $N\rightarrow e^{+}e^{-}\nu_{e}$ or 
ii) $N\rightarrow e^\pm \mu^\mp \nu_\mu$.
These modes lead to signatures with i) two separated electromagnetic showers and ii) one electromagnetic shower and one $\mu$-track, originating in the decay region.
As the experimental signatures require at least two charged leptons from $N$ decays, we do not recast the CHARM search for the single-$N_R$ scenarios.
However, the search is directly applicable to the scenario in which HNLs are produced through pair-$N_R$ effective operators and subsequently decay via standard mixing with the active neutrinos, as well as the ALP scenario where the ALP decays to a pair of charged leptons.

\item \textsc{\textbf{Belle:}} HNLs can also be searched for in leptonic and semileptonic $B$-meson decays. 
The Belle experiment, which was operated at the KEKB $e^{+}e^{-}$ collider mainly at the 
center-of-mass energy of 10.58~GeV (mass of the $\Upsilon(4S)$ resonance), searched for HNLs mixed in the electron sector through semi-leptonic two-body decays $N\rightarrow e^{\pm}\pi^{\mp}$~\cite{Belle:2013ytx}. 
All two- and three-body decays of $B$ mesons ($B^0$ and $B^\pm$)\footnote{Belle runs at the $\Upsilon(4S)$ resonance which decays to $B^+ B^-$ with a branching ratio of 51.4\% and to $B^0 \overline{B^0}$ with a branching ratio of 48.6\%~\cite{ParticleDataGroup:2022pth}.} were considered.
The search relies on the identification of a (prompt) charged lepton, and of a pion and a `signal' lepton with opposite charges arising from a common displaced origin.
The maximum sensitivity achieved for an HNL mass around 2 GeV is $|V_{eN}|^2 \sim 3 \times 10^{-5}$.
As the analysis strategy relies on identification of a prompt lepton coming from the $B$-meson decay, limits from this search are not applicable to meson decays triggered by the pair-$N_R$ effective operators or by the quark-flavor-violating (QFV) ALP effective couplings in association with a lighter meson.

Belle also searched for the rare decays $B\rightarrow h\nu\bar{\nu}$ with $h = \pi^{0}$, $K^{0}_{S}$, $\pi^{+}$, $K^{+}$, $K^{\ast+}$, $K^{\ast0}$,
$\rho^{+}$, $\rho^{0}$ with full luminosity of 711~fb$^{-1}$~\cite{Belle:2017oht}. The search strategy includes the reconstruction of an accompanying $B$-meson decaying semi-leptonically 
(\textit{i.e.} $B$-tag).
An updated search at Belle II operating at the SuperKEKB for $B^+ \rightarrow K^{+}\nu\bar{\nu}$ was performed with a different tagging method that exploits the inclusive decay of the other $B$ meson in the $\Upsilon(4S) \rightarrow B^{+}B^{-}$ event~\cite{Belle-II:2021rof}. Nevertheless, this last search provides a less stringent bound compared to~\cite{Belle:2017oht}, owing to the lower search luminosity of 63~fb$^{-1}$. These existing upper limits on branching ratios will apply if the HNL decay products are not detected. 
We will consider them in the next section, when discussing bounds on the Wilson coefficients of interest.

\item \textbf{BaBar:} For completeness, the BaBar experiment at SLAC has also searched for rare decays $B^{+}\rightarrow K^{+}\nu\bar{\nu}$ and $B^{0}\rightarrow K^{0}\nu\bar{\nu}$~\cite{BaBar:2010oqg,BaBar:2013npw} with a similar strategy as Belle, by reconstructing a recoiled $B$-meson decaying semileptonically to $B\rightarrow D^{(*)}l\nu$.
BaBar has also placed a more stringent limit with an hadronic tag for $B^{+}\rightarrow K^{+}\nu\bar{\nu}$ in Ref.~\cite{BaBar:2010oqg} . However, we consider the Belle one in~\cite{Belle:2017oht} in our calculations in order to be conservative.
BaBar has also searched for a dark photon directly in events with large missing transverse momenta and a single photon~\cite{BaBar:2017tiz}. This search can also be recast into bounds on the ALP-electron coupling relevant to our ALP scenario, although the limits are weak compared to other current bounds, so we do not include them in numerical studies.

\end{itemize}

In the following sections, the existing HNL searches at CHARM~\cite{CHARM:1985nku} and Belle~\cite{Belle:2013ytx} are reinterpreted with the proposed method in the EFT with HNLs and that of ALPs. 
In Table~\ref{tab:ULBRs}, we summarize to which of the benchmark scenarios considered in Sec.~\ref{sec:models} each past search is sensitive. For completeness, we also list the current bounds on branching ratios of rare meson decays we will use. Current bounds on the ALP coupling $c_{ee}$ are also shown.

%%%%%%%%
\begin{table}[t]
\renewcommand{\arraystretch}{1.2}
\centering
 \begin{tabular}{| l | c | }
 \hline
 Past HNL search & Sensitivity to   \\
 \hline
\hline
CHARM~\cite{CHARM:1985nku} & pair-$N_{R}$: 2HNL-D1, 2HNL-D2 and ALP-D  \\

Belle~\cite{Belle:2013ytx} & single-$N_{R}$: 1HNL-B1, 1HNL-B2 \\

\hline
\hline

 ALP bounds & Sensitivity to   \\
\hline
\hline

 Supernovae~\cite{Ferreira:2022xlw} & $c_{ee}$: ALP-D, ALP-B  \\
 E137~\cite{Essig:2010gu}	& $c_{ee}$: ALP-D, ALP-B	\\
  \hline
\hline
 Decay & Limit on BR \\
  \hline
\hline

 $D^{0} \rightarrow$ inv.~\cite{Belle:2016qek} & $9.4 \times 10^{-5}$ \\
 $D^{0} \rightarrow \pi^{0}\nu\bar{\nu}$~\cite{BESIII:2021slf} & $2.1 \times 10^{-4}$    \\

 $B^{0} \rightarrow$ inv.~\cite{BaBar:2012yut} & $2.4 \times 10^{-5}$  \\
 $B^{0} \rightarrow \pi^{0}\nu\bar{\nu}$~\cite{Belle:2017oht} & $9.0 \times 10^{-6}$ \\

 $B^0 \to K^{0}_{S} \nu \overline{\nu}$~\cite{Belle:2017oht} &   $1.3 \times 10^{-5}$ \\
		
$B^+ \to K^+ \nu \overline{\nu}$~\cite{Belle:2017oht} &  $1.9 \times 10^{-5}$ \\
 \hline
 \end{tabular}
 \caption{Recast searches and their sensitivities to our benchmarks in Tables~\ref{tab:benchmarksNN}, \ref{tab:benchmarksN} and \ref{tab:benchmarksALP}, as well as upper limits on the branching ratios (BRs) of (semi-)invisible $D$- and $B$-meson decays that will provide strong constraints on the corresponding HNL Wilson coefficients and QFV ALP couplings. Current limits constraining $c_{ee}$ are also quoted for completeness. }
 \label{tab:ULBRs}
\end{table}
%%%%%%%%

% !TEX root = ../reinterpretation.tex
\subsection{Future LHC far detectors}\label{subsec:future_exp}

Rather new LLP far detectors are already approved at the LHC: MoEDAL-MAPP1~\cite{Pinfold:2019nqj} and FASER~\cite{Feng:2017uoz}.
Their follow-up programs are also proposed to be operated with different integrated luminosities at the high-luminosity LHC (MoEDAL-MAPP2~\cite{Pinfold:2019zwp} with $300$~fb$^{-1}$ and FASER2~\cite{FASER:2018eoc} with $3$~ab$^{-1}$). Other experimental proposals include MATHUSLA~\cite{Chou:2016lxi,Curtin:2018mvb,MATHUSLA:2020uve}, ANUBIS~\cite{Bauer:2019vqk}, CODEX-b~\cite{Gligorov:2017nwh}, and more recently FACET~\cite{Cerci:2021nlb}. All these proposed detectors would be sensitive to detecting light long-lived particles decaying $\mathcal{O}(1)-\mathcal{O}(100)$ m, depending on their corresponding distance from the interaction point (IP). As being more than several meters away from the primary proton-proton collisions ensures a very low background environment, usually when assessing phenomenological prospects at these experiments, the assumption of a background free experiment is made when deriving $95\%$ C.L. upper limits on the model parameter space (corresponding to $N_{S}=3$).

Sensitivity estimates at these detectors were provided in detail in Ref.~\cite{DeVries:2020jbs} for the minimal HNL scenario and for $N_{R}$-LEFT operators with one HNL. In addition, in Ref.~\cite{Beltran:2022ast} estimates for pair-$N_{R}$ operators were studied. All details on the Monte-Carlo simulations and computation of decay probabilities at each experiment can be found in Refs.~\cite{DeVries:2020jbs,Beltran:2022ast}. Nevertheless, we highlight here some key aspects of the simulation procedure.

Decay probabilities of each simulated HNL take into account the far detector geometry and the HNL kinematics.
These are estimated with the help of Pythia8~\cite{Bierlich:2022pfr,Sjostrand:2014zea}, which we use to generate the production of $D$- and $B$-mesons from proton-proton collisions at $\sqrt{s}=14$ TeV at the LHC.
The mesons then decay to different channels that contain either one or two HNLs.
The meson decay branching ratios are computed analytically and fed into our Monte-Carlo simulation to compute the projected number of signal events, $N_{S}$.
This quantity also depends on the expected acceptance rate at each far detector experiment (\textit{i.e.}~defined as the average $\langle P[N \text{decay}] \rangle$ in Eq.~(4.1) of \cite{Beltran:2022ast}) (labeled as $P[\text{decay}]$ in Eq.~\eqref{eq:decayprob}). The exponential decay probability of each HNL in a detector generally depends on its boost factor including the traveling direction, its proper lifetime, and the detector location and geometries. Specific formulas that encapsulate the complicated geometrical shapes of each detector are detailed in Refs.~\cite{DeVries:2020jbs,Beltran:2022ast},\footnote{In Ref.~\cite{Beltran:2022ast}, we add projections at FACET, which were not included in Ref.~\cite{DeVries:2020jbs}.} and we strictly follow their procedure in this work for obtaining the projected limits in Fig.~\ref{fig:minimal}.

% !TEX root = ../reinterpretation.tex
\section{Example models}\label{sec:models}

The minimal ``3+1'' scenario assumes the existence of one HNL, $N$, that mixes with the active neutrinos $\nu_\ell$, 
$\ell = e,\,\mu,\,\tau$. The interaction Lagrangian of interest 
reads
\begin{equation}
 \mathcal{L}_\mathrm{min} = - \frac{g}{\sqrt{2}}\, V_{\ell N}\, 
 \overline{\ell} \gamma^\mu P_L N\, W_\mu^\dagger 
 - \frac{g}{2 \cos\theta_W}\, U_{\ell i}^\ast V_{\ell N} \,
 \overline{\nu_i} \gamma^\mu P_L N\, Z_\mu + \text{h.c.}\,,
\end{equation}
where $g$ is the $SU(2)_L$ gauge coupling, 
$\theta_W$ is the weak mixing angle,
$V_{\ell N}$ is active-heavy mixing, 
whereas $U_{\ell i}$ is active-light mixing 
(the PMNS mixing matrix).
The sum over $\ell = e,\,\mu,\,\tau$ and $i=1,\,2,\,3$ is implied. 
Note that we have neglected the terms quadratic in $V_{\ell N}$.
Production and decays of the HNL are governed by 
its mass $m_N$ and the mixing angles $V_{\ell N}$. 
In phenomenological studies, it is often assumed that the HNL mixes with one active flavor at a time. 
In this case, the model is characterized just by two independent parameters, mass and mixing $V_{\ell N}$. 
For this work, we focus on the mixing with the electron neutrino only.
Most of the searches for HNLs are interpreted in the framework of this simple model. 
For a generalization to the case of several HNLs, see \textit{e.g.} Refs.~\cite{Abada:2018sfh,Abada:2022wvh}.
Using the method formulated in Sec.~\ref{sec:method},
we will translate the bounds on the parameter space of the minimal scenario to the constraints on other (non-minimal) scenarios. 
We will consider two examples featured with long-lived fermions and (pseudo-)scalars, respectively: 
(i) the EFT with HNLs and (ii) ALPs.

\begin{figure}[t]
	\centering
	\includegraphics[width=0.48\textwidth]{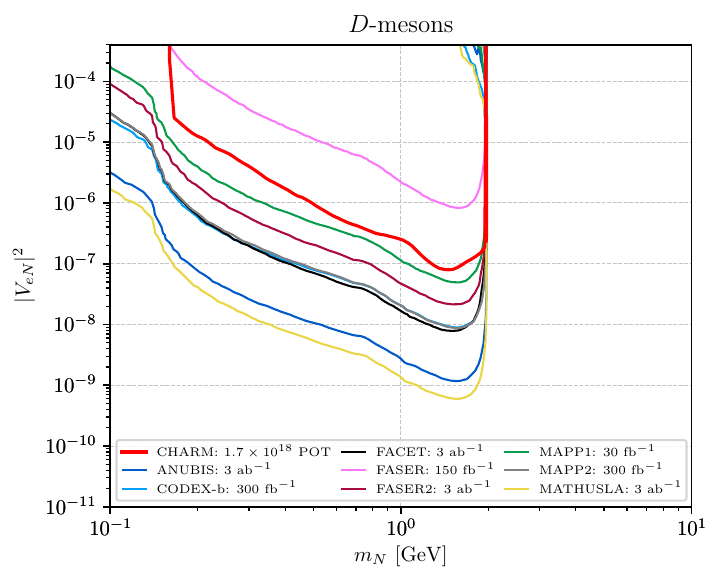}
	\includegraphics[width=0.48\textwidth]{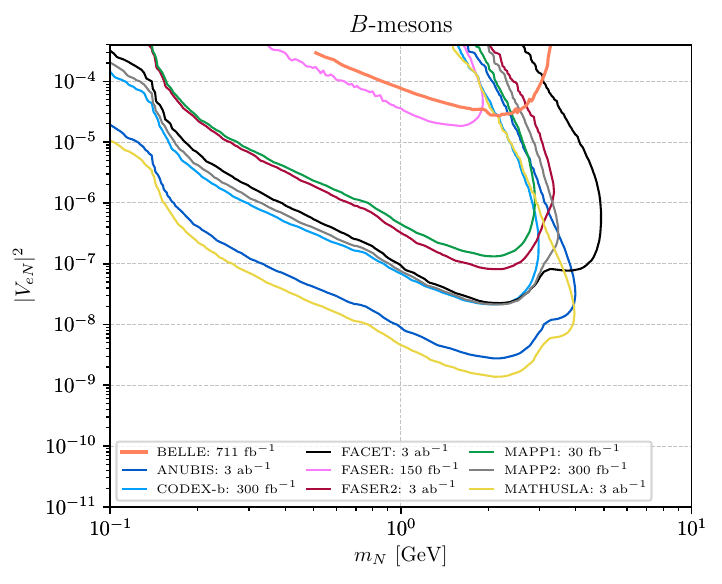}
	\caption{Sensitivity limits of LHC far detectors for HNLs produced from $D$-mesons (left) and $B$-mesons (right), respectively.
		We also show existing bounds on the minimal model obtained by CHARM~\cite{CHARM:1985nku} (left plot, red line) and Belle~\cite{Belle:2013ytx} (right plot, orange line).
	}
	\label{fig:minimal}
\end{figure}

In Fig.~\ref{fig:minimal}, we present bounds on the HNLs in the minimal scenario which dominantly mix with the electron neutrino, shown in the plane $|V_{eN}|^2$ vs.~$m_N$.
The left and right plots contain results for the HNLs produced from rare decays of $D$- and $B$- mesons, respectively.
In both plots, the exclusion limits of the LHC far detectors are at 95\% confidence level for 3 signal events with vanishing background, obtained by MC simulation following the procedures given in Refs.~\cite{DeVries:2020jbs,Beltran:2022ast}.
In the right plot, we have included results from the Belle experiment~\cite{Belle:2013ytx} with 711 fb$^{-1}$ integrated luminosity, and in the left, the results are overlapped with the existing bounds obtained at CHARM~\cite{CHARM:1985nku}, since these two experiments gave the leading bounds on the minimal HNLs produced from $B$- and $D$-mesons' decays, respectively.

%%%%%%%%
%%%%%%%%

For the far detectors, as we have performed MC simulations, we have three-dimensional data sets of $(m_N, |V_{eN}|^2, N_S)$, across the whole kinematically allowed mass range.
For this 3D dataset, we fix $|V_{eN}|^2$ to a small value such as $10^{-9}$, in order to ensure working in the large decay length limit.
However, for the existing bounds from Belle and CHARM, we do not have this full simulation information and can hence only make use of the published exclusion limits directly.
These datasets then allow us to perform reinterpretation into other models, such as the HNLs in the EFT, and ALPs, which are produced from bottom or charm meson decays.

% !TEX root = ../reinterpretation.tex
%%%%%%%%%%%%%%%%%%%%
\subsection{Effective field theory with HNLs}
%%%%%%%%%%%%%%%%%%%%
%
The SM extended by HNLs with masses below or around the electroweak scale $v$ should be viewed as an EFT, assuming heavy new physics exists at a scale $\Lambda \gg v$. 
Such an EFT is known as the $N_R$SMEFT~\cite{delAguila:2008ir,Aparici:2009fh,Liao:2016qyd}.
In this work, we are interested in GeV-scale HNLs produced in meson decays, and the suitable EFT is the $N_R$LEFT~\cite{Bischer:2019ttk,Chala:2020vqp,Li:2020lba,Li:2020wxi}, in which the heavy SM degrees of freedom 
(namely, the top quark, the Higgs boson, and the $Z$ and $W$ gauge bosons) are not present. 
The Lagrangian of this EFT reads
\begin{equation}
 \mathcal{L}_{N_R\mathrm{LEFT}} = \mathcal{L}_\mathrm{ren} 
 + \sum_{d \geq 5} \sum_{i} c_i^{(d)} \mathcal{O}_i^{(d)}\,, %
\end{equation}
where $\mathcal{L}_\mathrm{ren}$ is the renormalizable part of the Lagrangian (see \textit{e.g.} Eq.~(2.1) in Ref.~\cite{Beltran:2022ast}), $c_i^{(d)}$ are the Wilson coefficients of higher-dimensional operators $\mathcal{O}_i^{(d)}$, and the second sum goes over all independent operators of a given mass dimension $d$. 
The dimensionful coefficients $c_i^{(d)}$ scale as $v^{4-d}$.

We will focus on $d=6$ four-fermion operators with two quarks and either two or one $N_R$, since these contact interactions can mediate meson decays into HNLs.
In Table~\ref{tab:opsNN}, we summarize the pair-$N_R$ operators, assuming one generation of HNLs. 
Each of these operators carries quark flavor indices $i$ and $j$.
The last column indicates the number of independent real parameters associated with each operator structure. 
We recall that in the $N_R$LEFT, there are three generations of down-type quarks and two generations of up-type quarks.
%%%%%%%%
\begin{table}[t]  
\renewcommand{\arraystretch}{1.3}
\centering
 \begin{tabular}[t]{|c|c|c|}
    \hline
    \multicolumn{3}{|c|}{LNC operators} \\
    \hline
    Name & Structure& \# params \\
    \hline
   ${\cal O}_{dN}^{V,RR}$ &
    $\left(\overline{d_R}\gamma_{\mu}d_R\right)\left(\overline{N_R}\gamma^{\mu}N_R\right)$ & 
     9 \\
    ${\cal O}_{uN}^{V,RR}$ &
    $\left(\overline{u_R}\gamma_{\mu}u_R\right)\left(\overline{N_R}\gamma^{\mu}N_R\right)$ &
    4 \\ 
    \hline
    ${\cal O}_{dN}^{V,LR}$ &
    $\left(\overline{d_L}\gamma_{\mu}d_L\right)\left(\overline{N_R}\gamma^{\mu}N_R\right)$ & 
    9 \\
    ${\cal O}_{uN}^{V,LR}$ &
    $\left(\overline{u_L}\gamma_{\mu}u_L\right)\left(\overline{N_R}\gamma^{\mu}N_R\right)$ & 
    4 \\
    \hline
 \end{tabular}
 \hfill
 \begin{tabular}[t]{|c|c|c|}
    \hline
    \multicolumn{3}{|c|}{LNV operators} \\
    \hline
    Name & Structure & \# params \\
    \hline
    ${\cal O}_{dN}^{S,RR}$ &
    $\left(\overline{d_L}d_R\right)\left(\overline{N_R^c}N_R\right)$ & 
    18 \\
    ${\cal O}_{uN}^{S,RR}$ &
    $\left(\overline{u_L}u_R\right)\left(\overline{N_R^c}N_R\right)$ &
    8 \\ 
    \hline
    ${\cal O}_{dN}^{S,LR}$ &
    $\left(\overline{d_R}d_L\right)\left(\overline{N_R^c}N_R\right)$ & 
     18 \\
     ${\cal O}_{uN}^{S,LR}$ &
    $\left(\overline{u_R}u_L\right)\left(\overline{N_R^c}N_R\right)$ & 
     8 \\
     \hline
 \end{tabular}
 \caption{Four-fermion operators in the $N_R$LEFT, involving two quarks and two $N_R$, assuming one generation of HNLs.
 The third column shows the number of independent real parameters  associated with the given operator structure. 
 The LNV operator structures require ``+h.c.''.}
 \label{tab:opsNN}
\end{table}
%%%%%%%%
%
In Table~\ref{tab:opsN}, we list the single-$N_R$ operators with two quarks and a charged lepton. 
In addition to two quark flavor indices $i$ and $j$, these operators also carry a lepton flavor index $\ell = e,\,\mu,\,\tau$.
Assuming one generation of $N_R$, each operator structure encodes 36 independent real parameters. 
%%%%%%%%
\begin{table}[t]  
\renewcommand{\arraystretch}{1.3}
\centering
 \begin{tabular}[t]{|c|c|}
    \hline
    \multicolumn{2}{|c|}{LNC operators} \\
    \hline
    Name & Structure \\
    \hline
   ${\cal O}_{udeN}^{V,RR}$ &
    $\left(\overline{u_R}\gamma_{\mu}d_R\right)\left(\overline{e_R}\gamma^{\mu}N_R\right)$ \\ 
    \hline
    ${\cal O}_{udeN}^{V,LR}$ &
    $\left(\overline{u_L}\gamma_{\mu}d_L\right)\left(\overline{e_R}\gamma^{\mu}N_R\right)$ \\
    \hline
    ${\cal O}_{udeN}^{S,RR}$ &
    $\left(\overline{u_L} d_R\right)\left(\overline{e_L} N_R\right)$ \\ 
    ${\cal O}_{udeN}^{T,RR}$ &
    $\left(\overline{u_L} \sigma_{\mu\nu}d_R\right)\left(\overline{e_L} \sigma^{\mu\nu} N_R\right)$ \\
    \hline 
    ${\cal O}_{udeN}^{S,LR}$ &
    $\left(\overline{u_R} d_L\right)\left(\overline{e_L} N_R\right)$ \\ 
    \hline
 \end{tabular}
 \hspace{0.5cm}
 \begin{tabular}[t]{|c|c|}
    \hline
    \multicolumn{2}{|c|}{LNV operators} \\
    \hline
    Name & Structure \\
    \hline
   ${\cal O}_{udeN}^{V,LL}$ &
    $\left(\overline{u_L}\gamma_{\mu}d_L\right)\left(\overline{e_L}\gamma^{\mu}N_R^c\right)$ \\
    \hline
    ${\cal O}_{udeN}^{V,RL}$ &
    $\left(\overline{u_R}\gamma_{\mu}d_R\right)\left(\overline{e_L}\gamma^{\mu}N_R^c\right)$ \\
    \hline
    ${\cal O}_{udeN}^{S,LL}$ &
    $\left(\overline{u_R} d_L\right)\left(\overline{e_R} N_R^c\right)$ \\ 
    ${\cal O}_{udeN}^{T,LL}$ &
    $\left(\overline{u_R} \sigma_{\mu\nu}d_L\right)\left(\overline{e_R} \sigma^{\mu\nu} N_R^c\right)$ \\ 
    \hline 
    ${\cal O}_{udeN}^{S,RL}$ &
    $\left(\overline{u_L} d_R\right)\left(\overline{e_R} N_R^c\right)$ \\ 
    \hline
 \end{tabular}
 \caption{Four-fermion operators in the $N_R$LEFT, involving two quarks, one charged lepton and one $N_R$.
 For one generation of HNLs, there are 36 independent real parameters associated with each operator structure. 
 All operator structures require ``+h.c.''.}
 \label{tab:opsN}
\end{table}
%%%%%%%%
%

%%%%%%%%%%%%%%%%%%%%
\subsubsection{Four-fermion pair-$N_R$ operators}
\label{sec:pairNR}
%%%%%%%%%%%%%%%%%%%%
%
While the pair-$N_R$ operators may enhance the production 
of HNLs in meson decays, they do not trigger HNL decays 
(for one generation of HNLs).
Under the assumption that no other non-renormalizable interaction 
is present, HNLs decay via mixing with active neutrinos. 
Such a setup has been recently investigated in detail in the context of future LLP detectors at the LHC~\cite{Beltran:2022ast}.
Similarly, here we assume this mixing is with the electron neutrinos only.
For the far detectors, we compute the HNL visible decay width as including all the channels except the fully invisible ones (tri-neutrino), and for the CHARM search, we take into account only the purely leptonic channels with two charged leptons.

Each of the pair-$N_R$ operators carries quark flavor indices $i$ and $j$, \textit{e.g.~}$\mathcal{O}_{qN,ij}^{V,RR} = (\overline{q_{iR}} \gamma_\mu q_{jR}) (N_R \gamma^\mu N_R)$, with $q = u$ or $d$. 
The operator $\mathcal{O}_{uN,12}^{V,RR}$ induces $D^0 \to NN$ as well as a series of three-body decays $D \to P/V N N$, with $P~(V)$ denoting a lighter pseudoscalar (vector) meson. 
Similarly, $\mathcal{O}_{dN,31}^{V,RR}$ ($\mathcal{O}_{dN,32}^{V,RR}$) triggers $B^0 \to NN$ ($B_s^0 \to NN$) and a number of $B \to P/V N N$ decays. 
The same applies for the scalar-type operators given in the right panel of Table~\ref{tab:opsNN} and the operators with different chiralities ($LR$). 
As an example, we will restrict ourselves to some of these operators. 
The chosen benchmarks are listed in Table~\ref{tab:benchmarksNN}.
%%%%%%%%
\begin{table}[t]  
\centering
 \begin{tabular}[t]{|c|c|c|l|l|}
  \hline
  Benchmark & $P_\mathrm{prod}^{ij}$ & $P_\mathrm{decay}$ 
  & Production modes & Decay modes \\
  \hline
  \hline
  \multirow{3}{*}{2HNL-D1} & 
  \multirow{3}{*}{$c_{uN,12}^{V,RR}$} & 
  \multirow{3}{*}{$V_{eN}$} & 
  $D \to N + N$ & 
  \multirow{6}{*}{\parbox{3.5cm}{All decay modes via active-heavy mixing, see \textit{e.g.} Ref.~\cite{Bondarenko:2018ptm}.}} \\
  & & & $D \to \pi + N + N$ & \\
 & & & $D \to \eta^{(\prime)} + N + N$ & \\
  \cline{1-3}
   \multirow{3}{*}{2HNL-D2} & 
   \multirow{3}{*}{$c_{uN,12}^{S,RR}$} & 
   \multirow{3}{*}{$V_{eN}$} & 
   $D \to \rho + N + N$ & \\
  & & & $D \to \omega + N + N$ & \\
  & & & $D_s \to K^{(\ast)} + N + N$ & \\
  \hline
  \hline
  \multirow{3}{*}{2HNL-B1} & 
  \multirow{3}{*}{$c_{dN,31}^{V,RR}$} & 
  \multirow{3}{*}{$V_{eN}$} & 
  $B \to N + N$ & 
  \multirow{6}{*}{\parbox{3.5cm}{All decay modes via active-heavy mixing, see \textit{e.g.} Ref.~\cite{Bondarenko:2018ptm}.}}\\
  & & & $B \to \pi + N + N$ & \\
  & & & $B \to \eta^{(\prime)} + N + N$ & \\
  \cline{1-3}
  \multirow{3}{*}{2HNL-B2} & 
  \multirow{3}{*}{$c_{dN,31}^{S,RR}$} & 
  \multirow{3}{*}{$V_{eN}$} &  
  $B \to \rho + N + N$ & \\
  & & & $B \to \omega + N + N$ & \\
  & & & $B_s \to K^{(\ast)} + N + N$ & \\
  \hline
 \end{tabular}
 \caption{Example benchmarks from Ref.~\cite{Beltran:2022ast} 
 for which HNL production is mediated by a pair-$N_R$ operator 
 with certain quark flavor indices, whereas HNL decays
 proceed via active-heavy mixing.} 
 \label{tab:benchmarksNN}
\end{table}
%%%%%%%%
%
For the computation of the corresponding branching ratios, we refer the reader to Ref.~\cite{Beltran:2022ast}.

%%%%%%%%
\begin{figure}[t]
	\centering
	\includegraphics[width=1\textwidth]{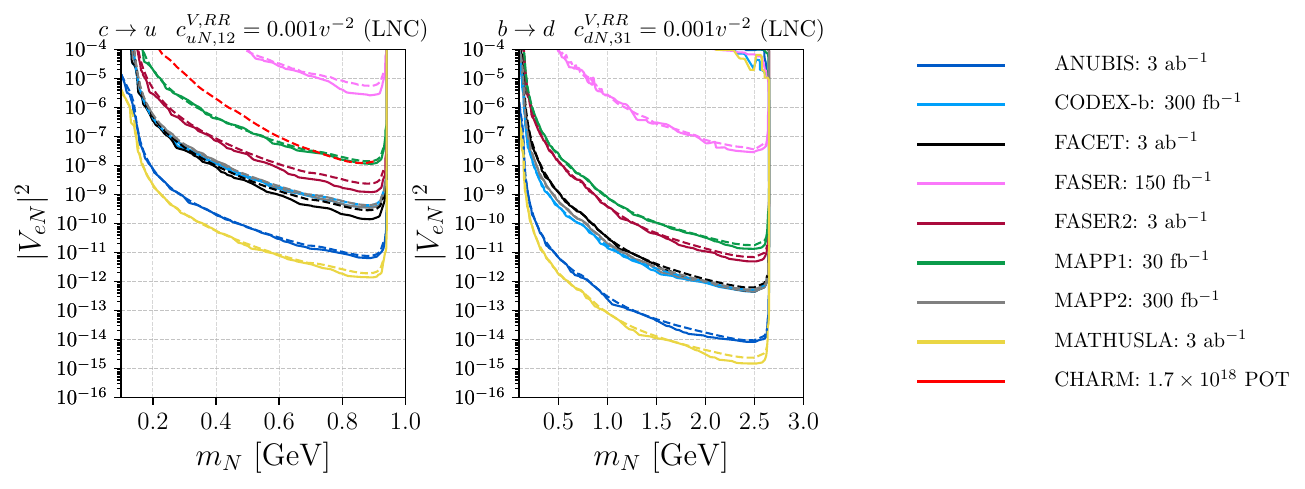}
	\includegraphics[width=1\textwidth]{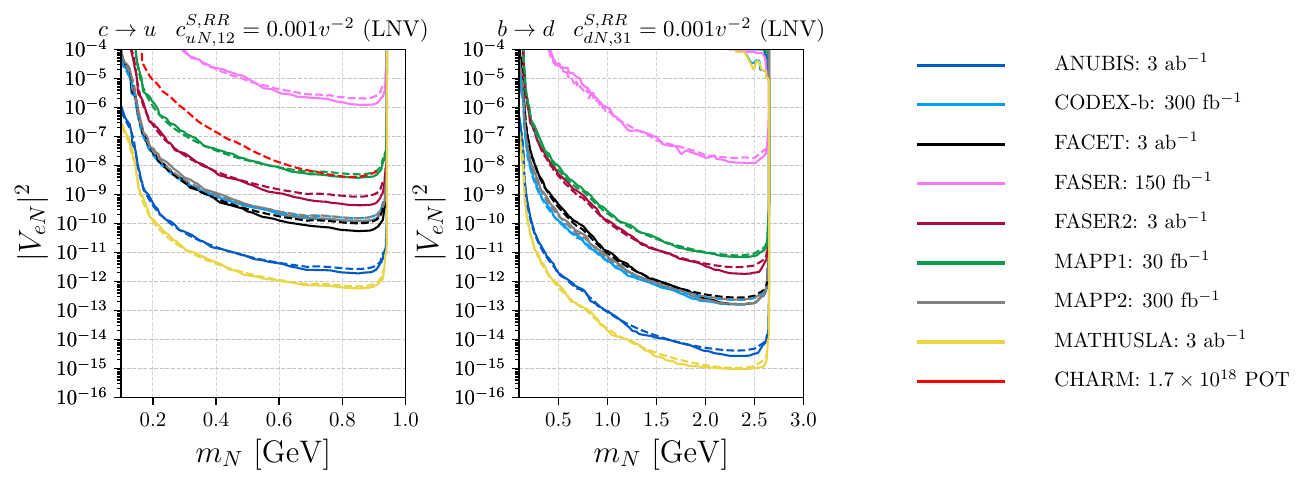}
	\caption{Reinterpretation results for the four selected pair-$N_R$ benchmarks, shown in the plane $|V_{eN}|^2$ vs.~$m_N$ for fixed corresponding Wilson coefficients at 0.001$v^{-2}$.
		The solid lines are simulation results obtained in Ref.~\cite{Beltran:2022ast} and dashed lines are reinterpretation results derived in the present work.
		The red dashed lines are for CHARM~\cite{CHARM:1985nku}.
	}
	\label{fig:doubleN_vv_vs_mass}
\end{figure}
%%%%%%%%

%%%%%%%%
\begin{figure}[t]
	\centering
	\includegraphics[width=1\textwidth]{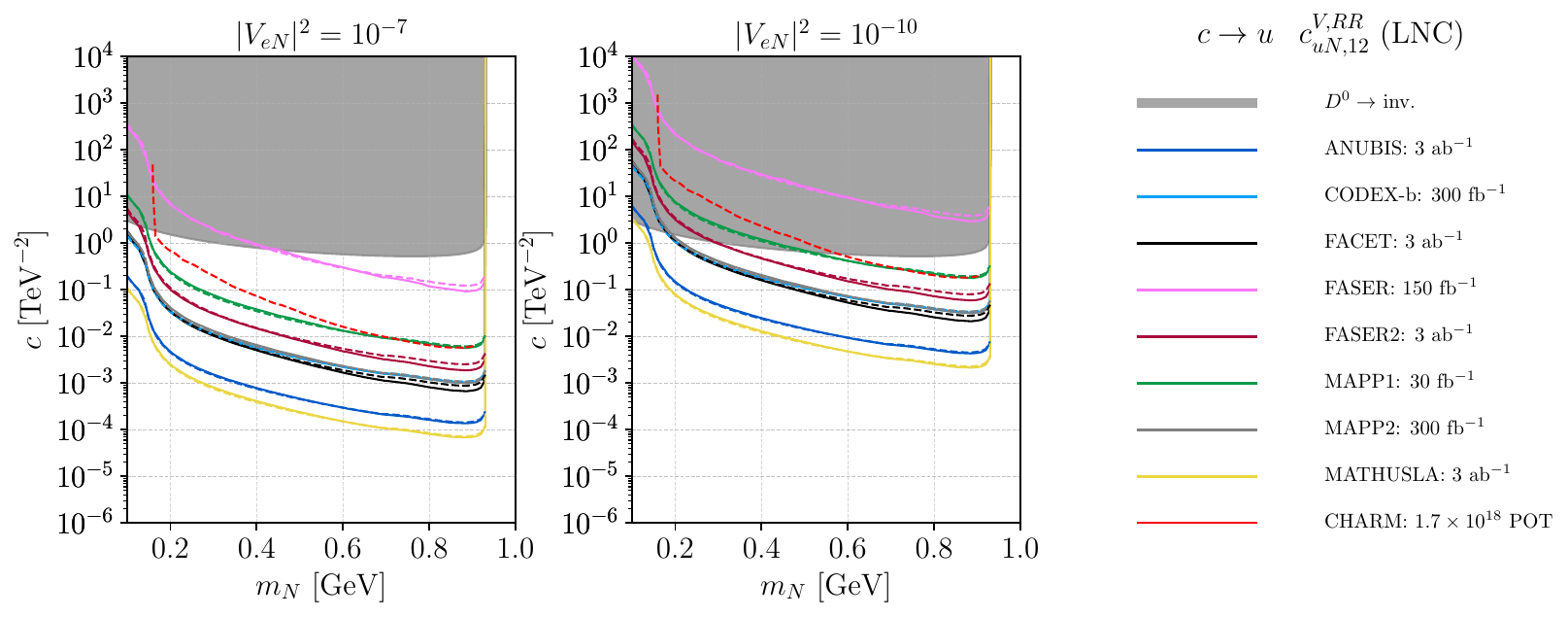}
	\includegraphics[width=1\textwidth]{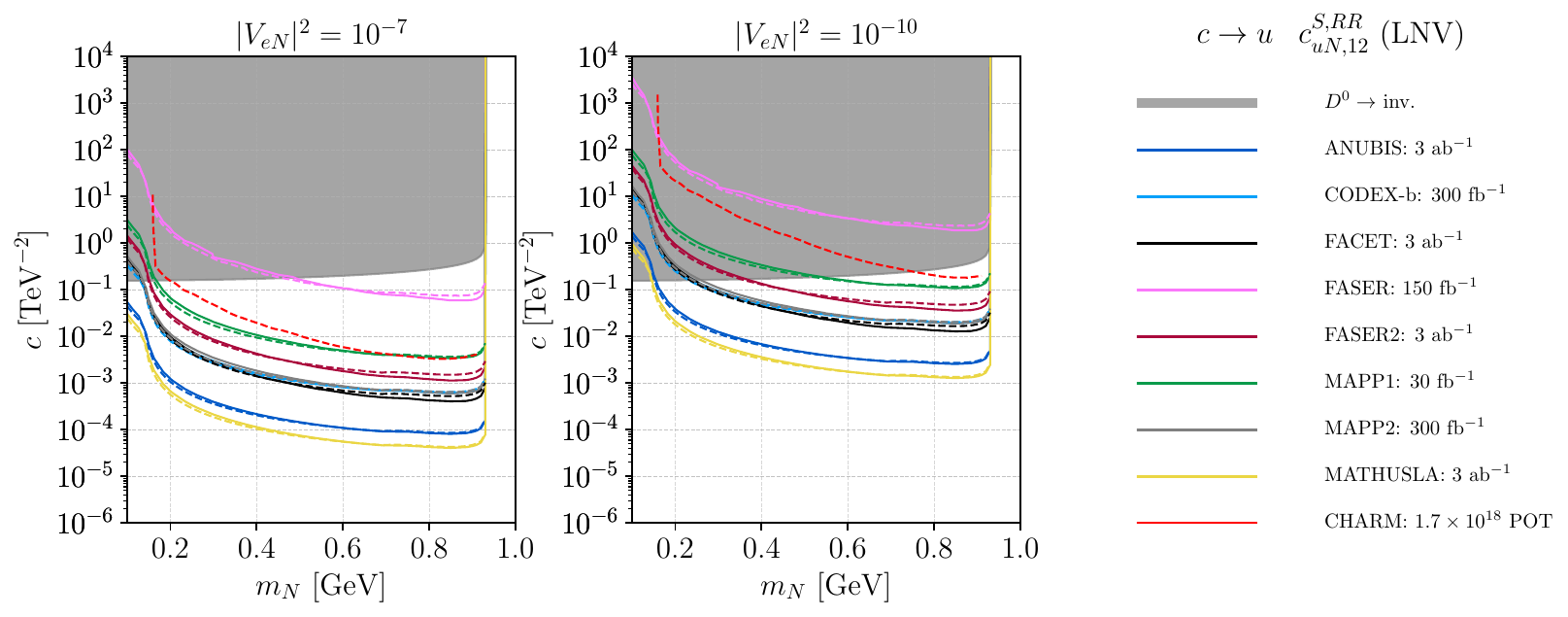}
	\caption{Reinterpretation results for the 2HNL-D1 (top panel) and 2HNL-D2 benchmarks (bottom panel), fixing $|V_{eN}|^2$ at $10^{-7}$ (left column) and $10^{-10}$ (right column), shown in the plane $c$ vs.~$m_N$.
		The solid lines are simulation results obtained in Ref.~\cite{Beltran:2022ast} and dashed lines are reinterpretation results derived in the present work.
		The red dashed lines are for CHARM~\cite{CHARM:1985nku}.
	}
	\label{fig:doubleN_coeff_vs_mass_12}
\end{figure}
%%%%%%%%

%%%%%%%%
\begin{figure}[t]
	\centering
	\includegraphics[width=1\textwidth]{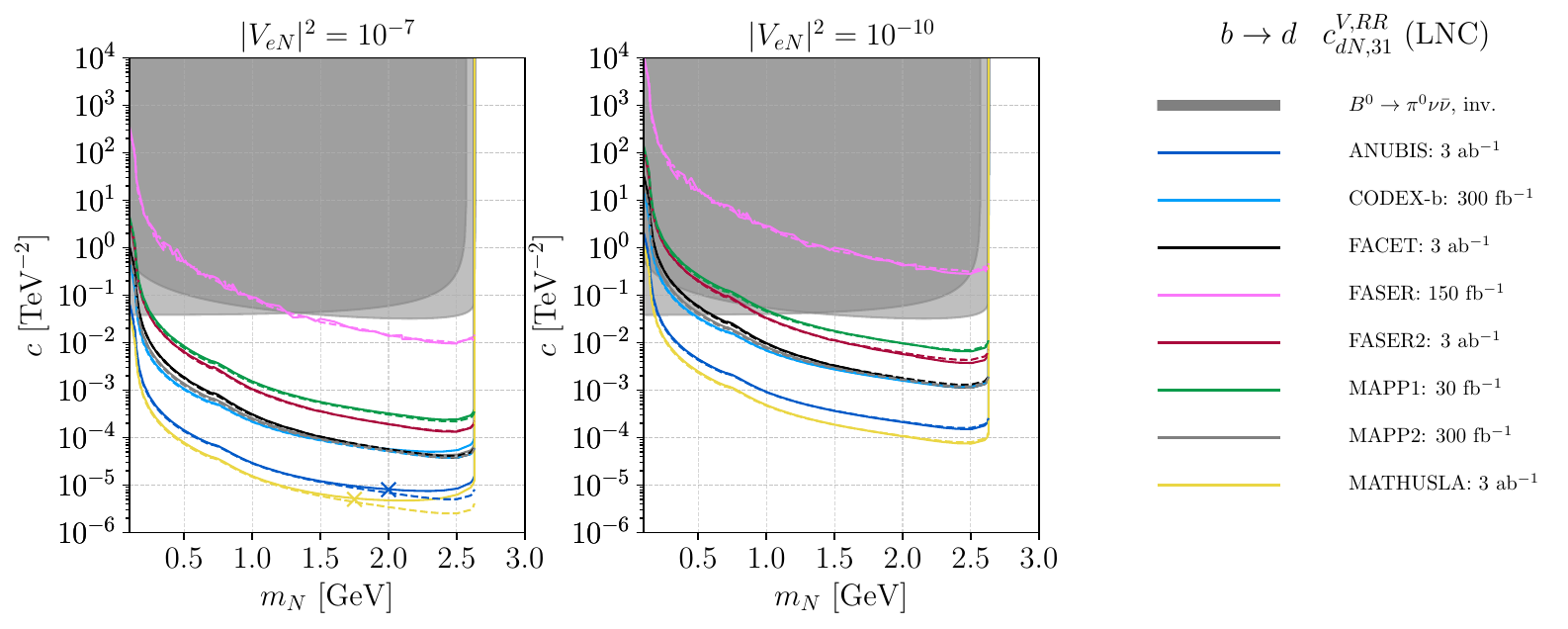}
	\includegraphics[width=1\textwidth]{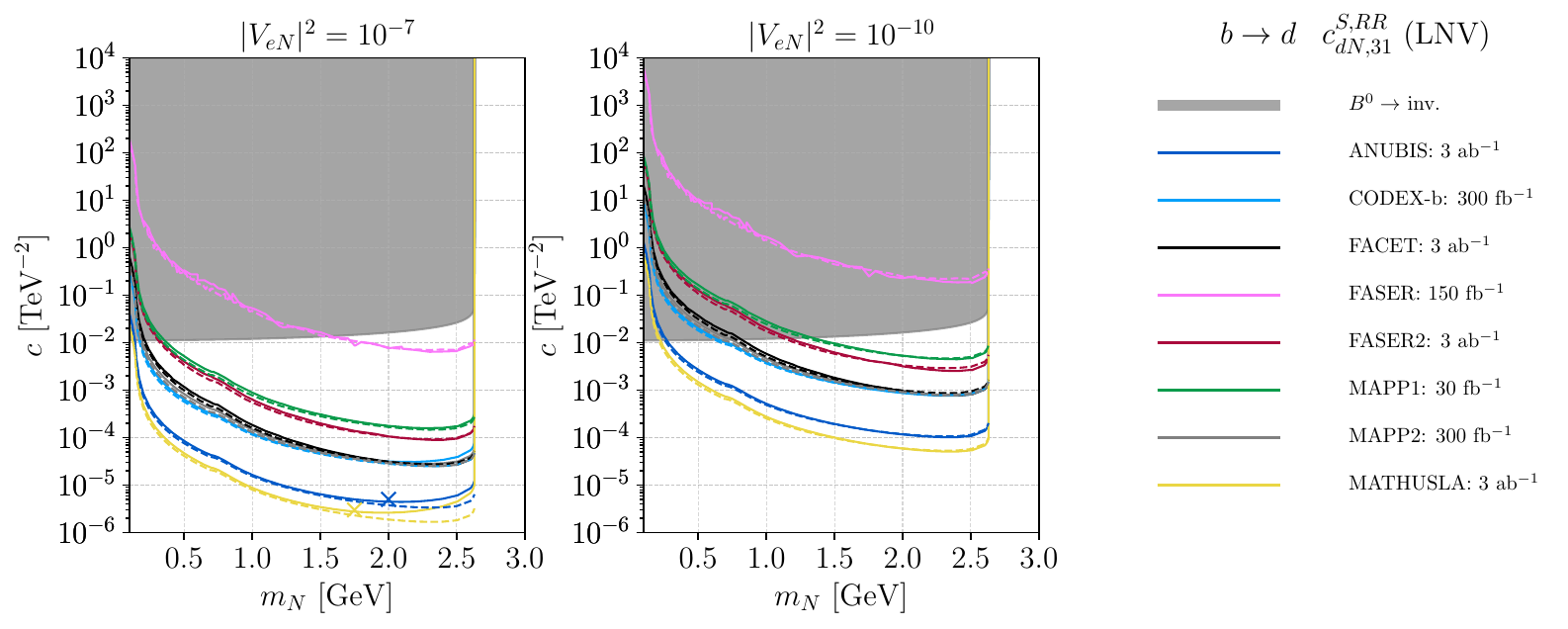}
	\caption{The same as Fig.~\ref{fig:doubleN_coeff_vs_mass_12} but for the 2HNL-B1 (top panel) and 2HNL-B2 benchmarks (bottom panel).
		The markers on the ANUBIS and MATHUSLA curves in the left plots correspond roughly to the positions where our recasting method's approximation starts to break down.
	}
	\label{fig:doubleN_coeff_vs_mass_31}
\end{figure}
%%%%%%%%

We show the reinterpretation results in Fig.~\ref{fig:doubleN_vv_vs_mass}, Fig.~\ref{fig:doubleN_coeff_vs_mass_12}, and Fig.~\ref{fig:doubleN_coeff_vs_mass_31}.
These are shown either in the $(m_N, |V_{eN}|^2)$ plane with the Wilson coefficients fixed at 0.001$v^{-2}$ (Fig.~\ref{fig:doubleN_vv_vs_mass}), or in the $(m_N,c)$ plane for $|V_{eN}|^2=10^{-7},10^{-10}$ (Fig.~\ref{fig:doubleN_coeff_vs_mass_12} and Fig.~\ref{fig:doubleN_coeff_vs_mass_31}) with $c$ denoting the Wilson coefficients.
For the far detectors, we extract the full simulation results (solid lines) from Ref.~\cite{Beltran:2022ast} and compare them with our reinterpreted ones (dashed lines).
Further, since the HNLs can decay to leptonic final states via mixing, the CHARM search~\cite{CHARM:1985nku} is sensitive to the two $D$-meson benchmarks.
On the other hand, as the Belle search~\cite{Belle:2013ytx} requires a prompt lepton, it is insensitive to the two $B$-meson benchmarks.
We find that for all the far detectors, the simulated and reinterpreted exclusion limits agree with each other very well in general, with very few exceptions.
For instance, in the plots in the left column of Fig.~\ref{fig:doubleN_coeff_vs_mass_31}, the reinterpreted bounds for ANUBIS and MATHUSLA are slightly too strong compared to the simulated ones.
This arises because for $|V_{eN}|^2=10^{-7}$ and $m_N\gtrsim 2.0$ GeV, the HNLs are not long-lived enough in the lab frame, compared to the distance of MATHUSLA from the CMS IP, impairing the approximation that our reinterpretation method takes to some extent.
For ANUBIS, similar behavior is observed, because of the relatively small pseudorapidity position of the detector and hence the small boost of the sterile neutrinos in the direction.
We place markers on the MATHUSLA and ANUBIS curves for $|V_{eN}|^2=10^{-7}$, roughly where our recasting method's approximation breaks down.
Moreover, for the two $D$-meson benchmarks, we find the reinterpreted CHARM search imposes exclusion limits comparable to those from MAPP1.
We stress here that our reinterpreted bounds are valid only for the large decay length limits, and hence do not work for the large $|V_{eN}|^2$ regime in Fig.~\ref{fig:doubleN_vv_vs_mass}.

Note that the invisible decay width constraint has been included in Fig.~\ref{fig:doubleN_coeff_vs_mass_12} and Fig.~\ref{fig:doubleN_coeff_vs_mass_31} (the gray area), implemented according to the relevant discussion in Sec.~\ref{subsec:past_exp}.

%%%%%%%%%%%%%%%%%%%%
\subsubsection{Four-fermion single-$N_R$ operators with a charged lepton}
\label{sec:singleNR}
%%%%%%%%%%%%%%%%%%%%
%
In contrast to the pair-$N_R$ operators, the same single-$N_R$ operator structure can mediate both HNL production and decay if two Wilson coefficients with different quark flavor indices are simultaneously present. 
Such a scenario for some of the operators given in Table~\ref{tab:opsN} has been studied in Ref.~\cite{DeVries:2020jbs} in the context of future far detectors at the LHC.
Namely, the operators that arise from $d = 6$ single-$N_R$ operators in the $N_R$SMEFT have been considered.
In the notation of Ref.~\cite{DeVries:2020jbs}, we have (see Eqs.~(8), (18) and (19) therein):
\begin{align}
 v^2 c_{udeN}^{V,RR} &= \overline{c}_{\rm VR}^{(6)} \approx C_{\rm VRR}^{(6)}\,, 
&&
 v^2 c_{udeN}^{V,LR} = \overline{c}_{\rm VL}^{(6)} \approx C_{\rm VLR}^{(6)}\,, \\
 v^2 c_{udeN}^{S,RR} &= \overline{c}_{\rm SR}^{(6)} \approx C_{\rm SRR}^{(6)}\,, 
&&
 v^2 c_{udeN}^{T,RR} = \overline{c}_{\rm T}^{(6)} \approx C_{\rm TRR}^{(6)}\,, \\
 v^2 c_{udeN}^{S,LR} &= \overline{c}_{\rm SL}^{(6)} \approx C_{\rm SLR}^{(6)}\,,
&&
 v^2 c_{udeN}^{V,LL} = \overline{C}_{\rm VL}^{(6)} \approx C_{\rm VLL}^{(6)}\,,
\end{align}
where the approximate equalities hold in the limit of negligible active-heavy mixing. 
Following the matching conditions given in Eq.~(11) of this reference, the VLR and VLL operators arise from 
the fermion-boson operators in the $N_R$SMEFT, whereas the remaining four operators are generated by the $N_R$SMEFT four-fermion operators.

In Ref.~\cite{DeVries:2020jbs}, two scenarios have been considered: (i) a leptoquark (LQ) model leading to $C_{\rm SRR}^{(6)} = 4 C_{\rm TRR}^{(6)}$ and (ii) a scenario  motivated by left-right symmetric models for which $C_{\rm VLR}^{(6)} \neq 0$.
Several benchmarks depending on the quark flavor indices of the operators responsible for HNL production and decay have been analyzed.
As an example, we choose benchmarks 1 and 3 from Ref.~\cite{DeVries:2020jbs}.
We rename benchmark 1.2 (1.3) as 1HNL-D1 (1HNL-D2)
and benchmark 3.2 (3.3) as 1HNL-B1 (1HNL-B2).
They are shown in Table~\ref{tab:benchmarksN}.
%%%%%%%%
\begin{table}[t]  
\centering
 \begin{tabular}[t]{|c|c|c|l|l|}
  \hline
  Benchmark & $P_\mathrm{prod}^{ij}$ & $P_\mathrm{decay}^{kl}$ 
  & Production modes & Decay modes \\
  \hline
  \hline
  \multirow{2}{*}{1HNL-D1} & 
  \multirow{2}{*}{$C_{\rm SRR}^{21} = 4 C_{\rm TRR}^{21}$} & 
  \multirow{2}{*}{$C_{\rm SRR}^{11} = 4 C_{\rm TRR}^{11}$} &
  $D \to e + N$ & \\
  & & & $D \to \pi + e + N$ & $N \to \pi + e$ \\
  \cline{1-3}
  \multirow{2}{*}{1HNL-D2} & 
  \multirow{2}{*}{$C_{\rm VLR}^{21}$} & 
  \multirow{2}{*}{$C_{\rm VLR}^{11}$} & 
  $D \to \rho + e + N$ & 
  $N \to \rho + e$ \\
  & & & $D_s \to K^{(\ast)} + e + N$ & \\
  \hline
  \hline
  \multirow{2}{*}{1HNL-B1} & 
  \multirow{2}{*}{$C_{\rm SRR}^{13} = 4 C_{\rm TRR}^{13}$} & 
  \multirow{2}{*}{$C_{\rm SRR}^{11} = 4 C_{\rm TRR}^{11}$} & 
  $B \to e + N$ & \\
  & & & $B \to \pi + e + N$ & $N \to \pi + e$ \\
  \cline{1-3}
  \multirow{2}{*}{1HNL-B2} & 
  \multirow{2}{*}{$C_{\rm VLR}^{13}$} & 
  \multirow{2}{*}{$C_{\rm VLR}^{11}$} & 
  $B \to \rho + e + N$ & 
  $N \to \rho + e$ \\
  & & & $B_s \to K^{(\ast)} + e + N$ & \\
  \hline
 \end{tabular}
 \caption{Example benchmarks from Ref.~\cite{DeVries:2020jbs} for which HNL production and decays are mediated by the same single-$N_R$ operator structure, but with different quark flavor indices for production and decay: $(ij) \neq (kl)$. 
 We rename benchmark 1.2 (1.3) in Ref.~\cite{DeVries:2020jbs} as 1HNL-D1 (1HNL-D2) and benchmark 3.2 (3.3) as 1HNL-B1 (1HNL-B2).}
 \label{tab:benchmarksN}
\end{table}
%%%%%%%%
%
For the computation of the corresponding branching ratios of meson and HNL decays, we refer the reader to Ref.~\cite{DeVries:2020jbs}.

\begin{figure}[t]
	\centering
	\includegraphics[width=1\textwidth]{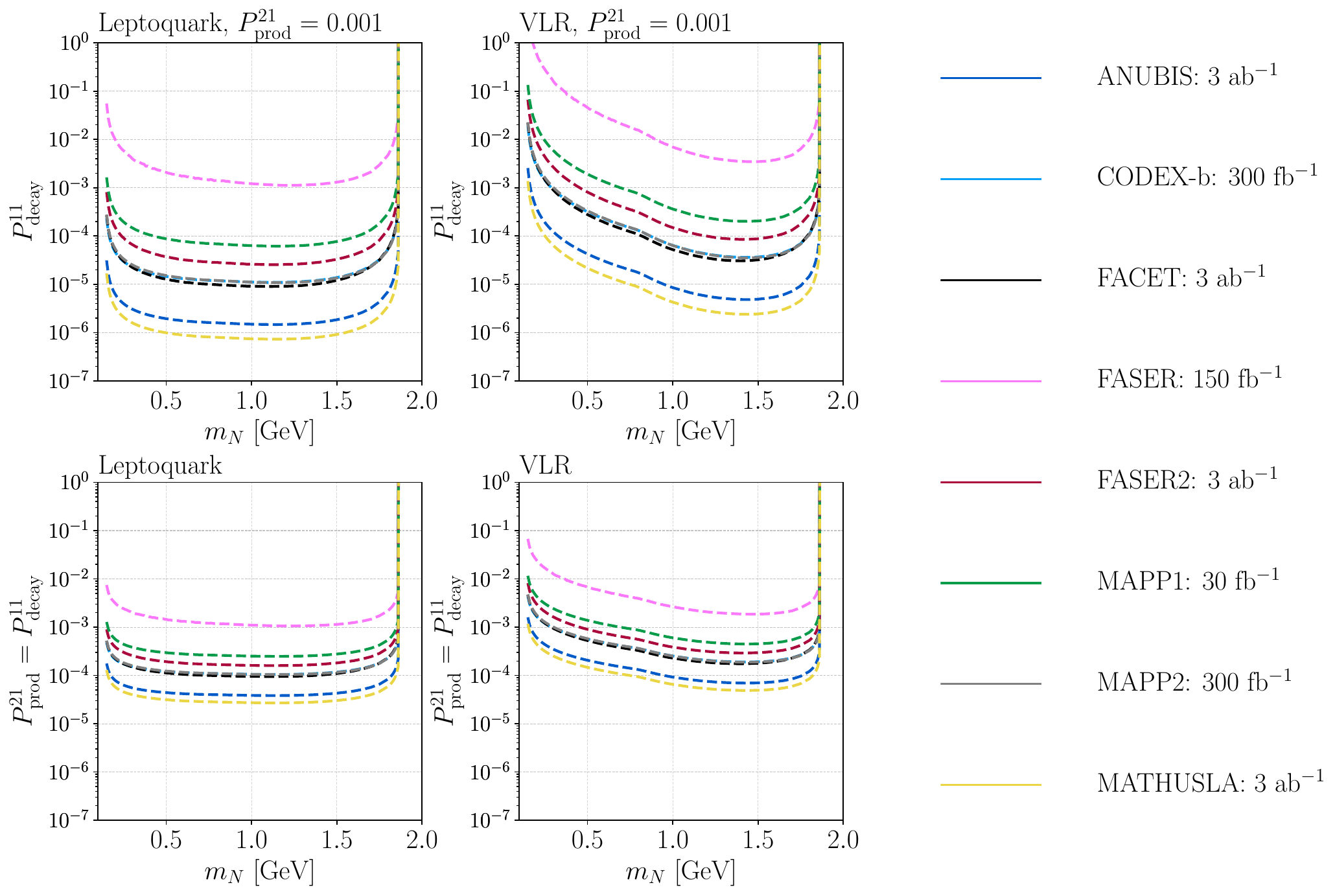}
	\caption{Reinterpretation results in dashed curves for benchmarks 1HNL-D1 (left panel) and 1HNL-D2 (right panel), shown in two types of parameter planes.
		The full simulation results obtained in Ref.~\cite{DeVries:2020jbs} are not overlapped, because here we turn off completely the minimal mixings, unlike in Ref.~\cite{DeVries:2020jbs}, leading to discrepancies in certain parameter regions where the non-zero mixing plays an important role, \textit{e.g.}~at masses below the pion threshold.
		The CHARM search~\cite{CHARM:1985nku} is not sensitive to this scenario and is hence not reinterpreted.
		Our reinterpretation method does not apply in the prompt regime, and hence the corresponding bounds for large decay couplings are not available.
	}
	\label{fig:singleN_2111}
\end{figure}

\begin{figure}[t]
	\centering
	\includegraphics[width=1\textwidth]{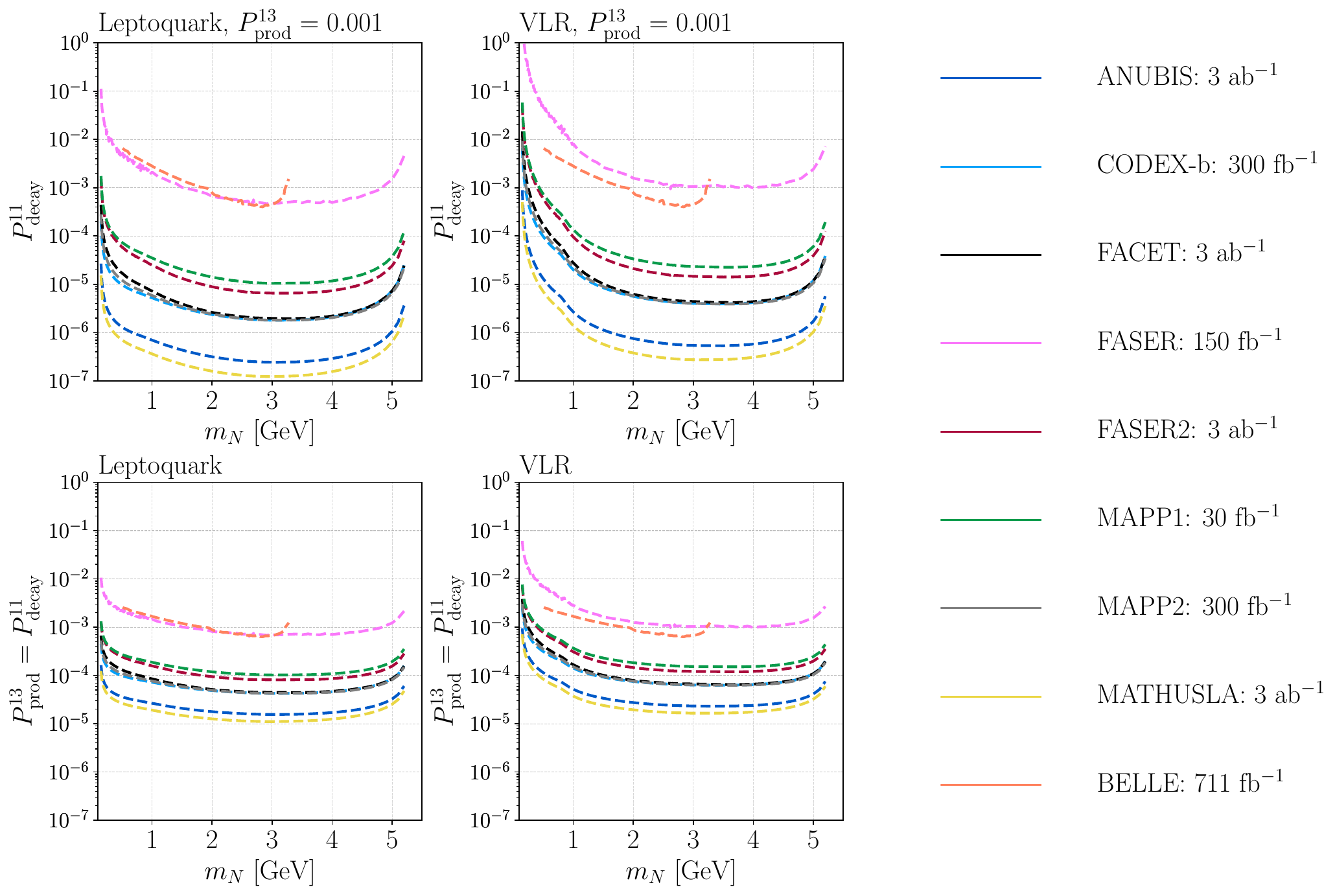}
	\caption{The same figure as Fig.~\ref{fig:singleN_2111}, but for benchmarks 1HNL-B1 and 1HNL-B2.
		In addition, here, besides the LHC far detectors, we also reinterpret the Belle search~\cite{Belle:2013ytx}.
		As in Fig.~\ref{fig:singleN_2111}, the ``prompt bounds'' are not derivable with the fast reinterpretation method and are hence not displayed.
	}
	\label{fig:singleN_1311}
\end{figure}

The reinterpretation results are given in Fig.~\ref{fig:singleN_2111} and Fig.~\ref{fig:singleN_1311}, for the two benchmark scenarios, respectively.
Here, we assume that the minimal mixing $|V_{eN}|^2$ is vanishing, and the production and decay of the HNLs are solely induced by the two single-$N_R$ operators switched on in each benchmark; this is somewhat different from the approach taken in Ref.~\cite{DeVries:2020jbs} where the full simulation method was used and the minimal mixing $|V_{eN}|^2$ was taken to be equal to $m_{\nu}/m_N=(0.05 \text{ eV})/m_N$ as a representative value leading to the HNL participation in weak interaction with the mixing.
Since we assume vanishing $|V_{eN}|^2$, our results show discrepancies from those shown in Ref.~\cite{DeVries:2020jbs} in certain parameter regions.
For example, our curves have no sensitivity below the pion mass threshold.
Therefore, we only show the reinterpreted exclusion limits (with dashed lines), but not the full simulated results; comparison can nevertheless be performed by cross-checking various parameter points against the plots given in Ref.~\cite{DeVries:2020jbs}.
In Fig.~\ref{fig:singleN_2111} containing results for HNLs produced from $D$-meson decays, we consider only the LHC far detectors since the CHARM search~\cite{CHARM:1985nku} requires leptonic final states while the HNLs in this benchmark decay to $e^\pm \pi^\mp$, where the charged pion was not searched for in Ref.~\cite{CHARM:1985nku}.
However, for the $B$-meson benchmarks (1HNL-B1 and 1HNL-B2), the Belle search~\cite{Belle:2013ytx} can be recast into bounds, as these benchmarks have a prompt electron from $B$-decays and their final states were also searched for in Ref.~\cite{Belle:2013ytx}.
Here, we take into account both $B^\pm$ and $B^0$ meson contributions.
These new results are shown in Fig.~\ref{fig:singleN_1311} (orange for Belle).
We observe that the reinterpreted Belle bounds are rather weak and only comparable to those of FASER in the plotted mass range;\footnote{In principle, if the production and decay EFT couplings are decoupled as in the case of the upper plots of Fig.~\ref{fig:singleN_1311}, these experiments including Belle should be sensitive to the whole kinematically allowed mass range. However, since without a simulation for the minimal-scenario search at Belle we can only make use of the published exclusion limits for the reinterpretation, our method gives results covering only the mass range shown in the minimal-scenario bounds.} all the other far detectors can probe new parameter space.

In general, for the parameter regions where the effect from the non-zero mixing angle is negligible, we find excellent agreement between our reinterpreted exclusion limits and the results given in Ref.~\cite{DeVries:2020jbs}, with a single exception: in the lower plots of Fig.~\ref{fig:singleN_1311} shown in the plane $P^{13}_{\text{prod}}=P^{11}_{\text{decay}}$ vs.~$m_N$, our reinterpretation results for FASER are sensitive across the whole kinematically allowed mass range, while the corresponding curves worked out with a full simulation in Ref.~\cite{DeVries:2020jbs} have a slightly smaller upper mass reach; we over-estimate the sensitivities to some extent with the reinterpretation method.
Indeed, this can happen if the production and decay couplings are related.
However, in the case that they are decoupled, the whole kinematically allowed mass range is usually covered by the exclusion bounds, and this issue would not arise.

Before we close this subsection, we shall mention again that our reinterpretation only works for long decay length, and therefore, for instance, in Fig.~\ref{fig:singleN_2111} and Fig.~\ref{fig:singleN_1311}, there are only lower parts of the sensitivity curves and the upper parts are absent; as the upper curves would correspond to the promptly decaying regime.

%%%%%%%%
%%%%%%%%

%%%%%%%%
%%%%%%%%

% !TEX root = ../reinterpretation.tex
%%%%%%%%%%%%%%%%%%%%
\subsection{Axion-like particles}
%%%%%%%%%%%%%%%%%%%%
%
Another example, which may feature LLPs produced in meson decays, is the low-energy EFT of ALPs~\cite{Georgi:1986df,Choi:1986zw}. 
This theory has received significant attention in recent years (see \textit{e.g.}~Refs.~\cite{Brivio:2017ije,Chala:2020wvs,Bauer:2020jbp,Bauer:2021mvw} and the references therein).
The low-energy Lagrangian for the ALP $a$ to $d=5$ reads
\begin{equation}
 \mathcal{L}_\mathrm{ALP} = \frac{1}{2}\, \partial_\mu a\, \partial^\mu a 
 - \frac{1}{2}\, m_a^2\, a^2 
 + \partial_\mu a \left[
 \sum_q \sum_{i,j} c^q_{ij}\, \overline{q_i} \gamma^\mu q_j 
 +\sum_l \sum_{\ell,\ell'} c^l_{\ell\ell'}\, \overline{l_\ell} \gamma^\mu l_{\ell'}
 \right] + \dots\,,
\end{equation}
where $m_a$ denotes the ALP mass; $q$ runs over $u_L$, $u_R$, $d_L$, $d_R$; $l$ goes over $e_L$, $e_R$, $\nu_L$; and the dots stand for the couplings of the ALP to anomalous gauge currents. 
This Lagrangian is approximately invariant under the shift symmetry, $a \to a + \text{constant}$.\footnote{The shift symmetry is broken by the ALP mass term and the gauge anomalous couplings.}

We will focus on flavor off-diagonal ALP couplings to quarks and flavor-diagonal couplings to charged leptons, as a representative benchmark for our study. 
For the latter, only the coupling to the axial-vector current is relevant.\footnote{Upon integrating by parts and applying the equation of motion for a fermion field, the coupling to the vector current reduces to a total derivative.}
Namely, for charged leptons $\ell = e$, $\mu$, $\tau$, we have
\begin{equation}
 \partial_\mu a \left[c^{e_L}_{\ell\ell}\, \overline{\ell} \gamma^\mu P_L \ell + c^{e_R}_{\ell\ell}\, \overline{\ell} \gamma^\mu P_R \ell \right] 
 \to
 \frac{c_{\ell\ell}}{2}\, \partial_\mu a\, \overline{\ell} \gamma^\mu \gamma_5 \ell
 \quad \text{with} \quad
 c_{\ell\ell} = c^{e_R}_{\ell\ell} - c^{e_L}_{\ell\ell}\,.
\end{equation}
For $m_a > 2 m_\ell$, this coupling triggers the decay $a \to \ell \bar{\ell}$ with the decay rate~\cite{Bauer:2017ris}
\begin{equation}
 \Gamma\left(a \to \ell^+ \ell^-\right) = \frac{c_{\ell\ell}^2}{8\pi}\, m_a m_\ell^2\, \sqrt{1 - \frac{4 m_\ell^2}{m_a^2}}\,.
\end{equation}

Concerning the quark flavor indices, we will consider two possibilities: 
(i) up-type quarks with $i=1$ and $j=2$ leading to the $c \to u$ transitions, \textit{i.e.}~$D \to (\pi\,,~\eta^{(\prime)}\,,~\rho\,,~\omega) + a$ and $D_s^+ \to K^{(\ast)+} + a$, and 
(ii) down-type quarks with $i=3$ and $j=2$ realizing the $b \to s$ transitions, \textit{i.e.}~$B \to K^{(\ast)} +a$ and $B_s^0 \to (\eta^{(\prime)}\,,~\phi) + a$. 
The corresponding decay rates read~\cite{Bauer:2021mvw}:
\begin{align}
 \Gamma\left(P \to P^{\prime} a\right) &= f\, \frac{|c^q_{ij}|^2}{64\pi}
 \left|F_0^{P \to P'}(m_a^2)\right|^2 m_P^3 \left(1 - \frac{m_{P'}^2}{m_P^2}\right)^2
 \lambda^{1/2}\left(\frac{m_{P'}^2}{m_P^2}, \frac{m_a^2}{m_P^2}\right)\,,\label{eq:PtoPprimea1} \\
 \Gamma\left(P \to V a\right) &= g\, \frac{|c^q_{ij}|^2}{64\pi} 
 \left|A_0^{P \to V}(m_a^2)\right|^2 m_P^3 \,  \lambda^{3/2}\left(\frac{m_{V}^2}{m_P^2}, \frac{m_a^2}{m_P^2}\right)\,, \label{eq:PtoV}
\end{align}
where $F_0^{P \to P'}$ and $A_0^{P \to V}$ are the corresponding form factors defined in Ref.~\cite{Wirbel:1985ji}, and
\begin{equation}
 \lambda(x,y) = 1 + x^2 + y^2 - 2x - 2y - 2xy\,.
\end{equation} 
The numerical factors $f$ and $g$ are different from 1 only in some transitions involving neutral mesons.
In particular, we have $f=1/2$ for  $D^0 \to \pi^0$, $f=2/3$ for $D^0\to \eta$ and $B_s^0 \to \eta $, and $f=1/3$ for $D^0 \to \eta^\prime$ and $B_s^0 \to \eta^\prime$, whereas $g=1/2$ for $D^0 \to \rho^0$ and $D^0 \to \omega$.
For $q=u$ and $(i,j)=(1,2)$, we have
\begin{align}
 c^u_{12} = c^{u_R}_{12} + c^{u_L}_{12}
 \qquad &\text{for} \qquad 
  D^+ \to \pi^+,~ D_s^+ \to K^+, \nonumber \\
 &\text{and} \qquad 
 %(P^0,P^{\prime0}) = (D^0,\pi^0)\,,~(D^0,\eta)\,,~(D^0,\eta^\prime)\,, \\ 
 D^0 \to \pi^0,~D^0 \to \eta,~D^0 \to \eta^\prime\,, \\ 
 c^u_{12} = c^{u_R}_{12} - c^{u_L}_{12}
 \qquad &\text{for} \qquad
 %(P^+,V^+) = (D^+,\rho^+)\,,~(D_s^+,K^{\ast+})\,, \nonumber \\
 D^+ \to \rho^+,~D_s^+ \to K^{\ast+}, \nonumber \\
 &\text{and} \qquad
 D^0 \to \rho^0,~ D^0 \to \omega\,.
\end{align}
For $q=d$ and $(i,j) = (3,2)$, we have instead,
\begin{align}
 c^d_{32} = c^{d_R}_{32} + c^{d_L}_{32}
 \qquad &\text{for} \qquad
 B^+ \to K^+, \nonumber \\
 &\text{and} \qquad
% (P^0,P^{\prime0}) = (B^0,K^0)\,,~(B_s^0,\eta)\,,~(B_s^0\,,\eta')\,, \\
B^0 \to K^0,~B_s^0 \to \eta,~B_s^0 \to \eta', \\
 c^d_{32} = c^{d_R}_{32} - c^{d_L}_{32}
 \qquad &\text{for} \qquad
 B^+ \to K^{\ast+}, \text{ and } 
% (P^0,V^0) = (B^0,K^{\ast0})\,,~(B_s^0,\phi)\,.
B^0 \to K^{\ast0},~ B_s^0 \to \phi\,.
\end{align}
In the numerical study, we will assume that either $c^{q_R}_{ij} = 0$ or $c^{q_L}_{ij} = 0$, such that a single coupling controls both $P \to P^\prime + a$ and $P \to V + a$ decays.
The considered benchmarks are shown in Table~\ref{tab:benchmarksALP}.
%%%%%%%%
\begin{table}[t]  
\centering
 \begin{tabular}[t]{|c|c|c|l|l|}
  \hline
  Benchmark & $P_\mathrm{prod}^{ij}$ & $P_\mathrm{decay}$ 
  & Production modes & Decay modes \\
  \hline
  \hline
  \multirow{5}{*}{ALP-D} & 
  \multirow{5}{*}{$c^u_{12}$} & 
  \multirow{5}{*}{$c_{ee}$} &
  $D \to \pi + a$ & 
  \multirow{5}{*}{$a \to e^+ + e^-$} \\
  & & & $D \to \eta^{(\prime)} + a$ & \\
  & & & $D \to \rho + a$ & \\
  & & & $D \to \omega + a$ & \\
  & & & $D_s \to K^{(\ast)} + a$ & \\
  \hline
  \hline
  \multirow{3}{*}{ALP-B} & 
  \multirow{3}{*}{$c^d_{32}$} & 
  \multirow{3}{*}{$c_{ee}$} & 
  $B \to K^{(\ast)} + a$ & 
  \multirow{3}{*}{$a \to e^+ + e^-$} \\
  & & & $B_s \to \eta^{(\prime)} + a$ & \\
  & & & $B_s \to \phi + a$ & \\  
  \hline
 \end{tabular}
 \caption{Example benchmarks for the scenario in which an ALP 
 is produced through a flavor off-diagonal coupling to quarks ($i\neq j$) 
 and subsequently decays via a flavor-diagonal coupling to charged leptons.
}
 \label{tab:benchmarksALP}
\end{table}
%%%%%%%%
%\
The charge-conjugated channels are included in the computation.

We take the transition form factors $F_0^{D\to P'}(q^2)$ from Ref.~\cite{Lubicz:2017syv} for $D\to \pi$ and $F_0^{D\to P'}(q^2)$ and $A_0^{D\to V}(q^2)$ from Ref.~\cite{Ivanov:2019nqd} for the other $D$-meson decays listed in Table~\ref{tab:benchmarksALP}.
For the $B$-meson decays, we use $F_0^{B\to P'}(q^2)$ from Ref.~\cite{Aoki:2021npn} for $B\to K$ and from Ref.~\cite{Wu:2006rd} for $B_s^0 \to \eta^{(\prime)}$; we extract $A_0^{B\to V}(q^2)$ from Ref.~\cite{Bharucha:2015bzk} for $B\to K^\ast$ and $B_s^0 \to \phi$.
In all these cases we evaluate the form factors at $q^2 = m_a^2$. 
For more details on the form factors used, 
we refer the reader to Appendix~A.3 of Ref.~\cite{Beltran:2022ast}.\footnote{We note that $A_0^{P \to V}$ used here is related to the form factors employed in Ref.~\cite{Beltran:2022ast} as
	$$\left|A_0^{P\to V}(q^2)\right| = \frac{1}{2 m_V} \left|f(q^2) + \left(m_P^2 - m_V^2\right) a_+(q^2) + q^2 a_-(q^2)\right|\,.$$}
\begin{figure}[t]
 \centering
 \includegraphics[height=0.29\textheight]{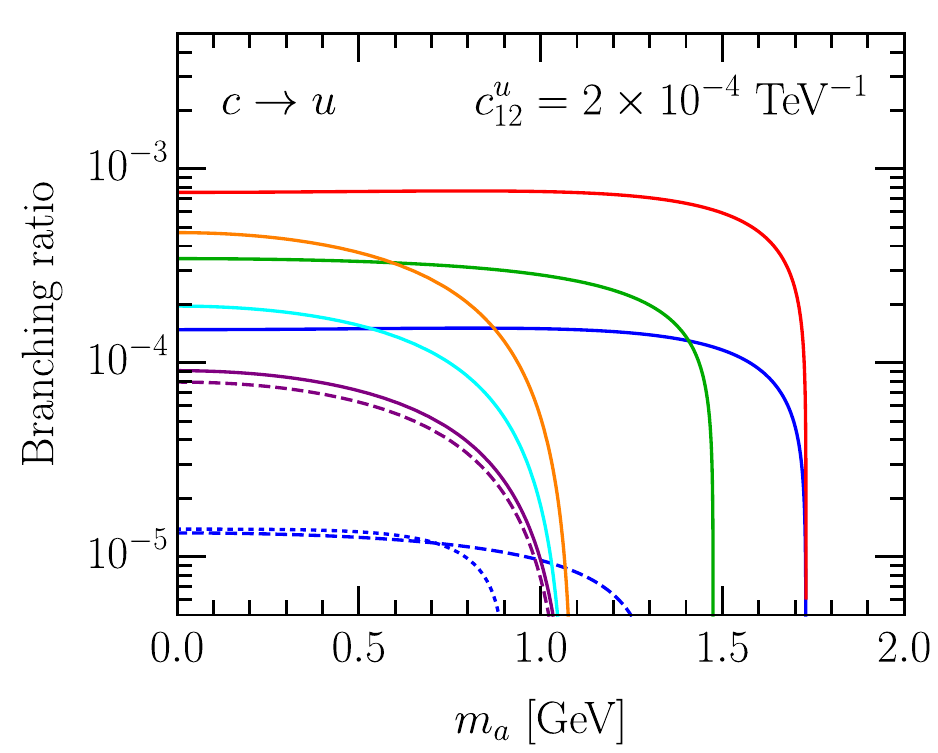}
 \includegraphics[height=0.29\textheight]{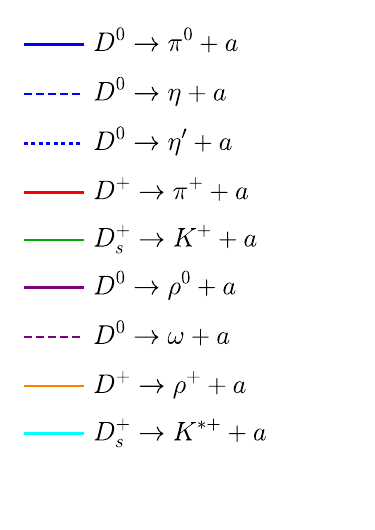}\\[0.15cm]
 \includegraphics[height=0.29\textheight]{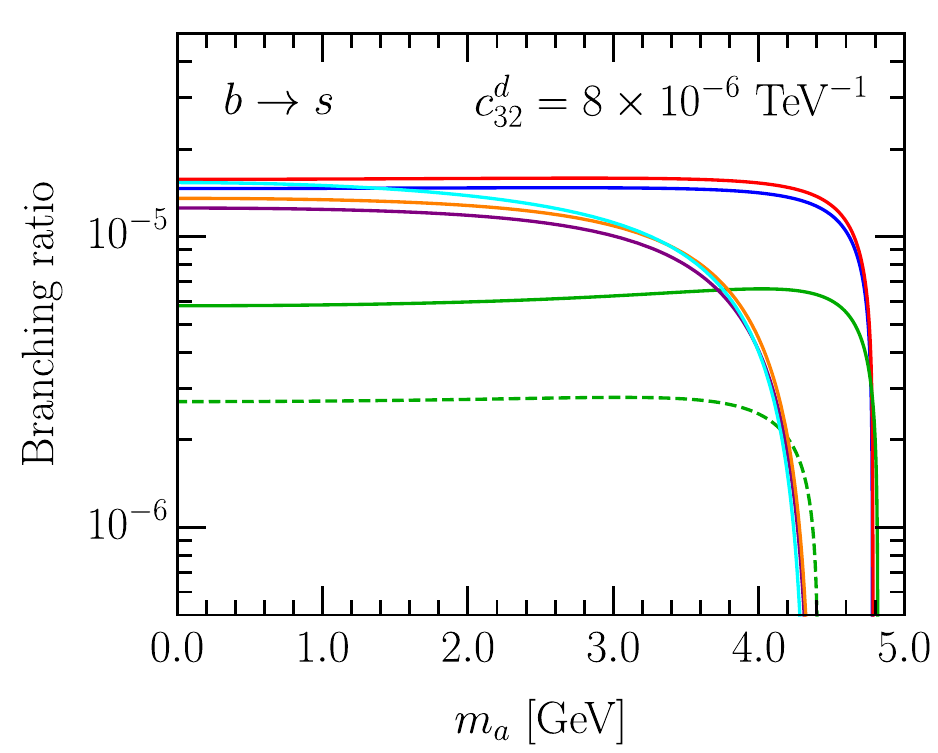}
 \includegraphics[height=0.29\textheight]{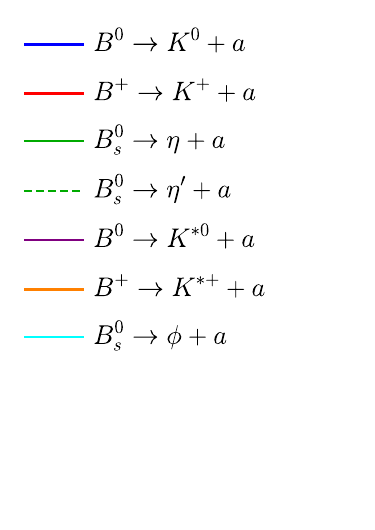}\\
 \caption{Branching ratios of $D$- and $B$-meson decays 
 to a lighter meson and the ALP triggered 
by the couplings $c^u_{12}$ and $c^d_{32}$, respectively. 
The couplings have been fixed as 
$c^u_{12} = 2 \times 10^{-4}$~TeV$^{-1}$ (top)
and $c^d_{32} = 8 \times 10^{-6}$~TeV$^{-1}$ (bottom).}
 \label{fig:BRs}
\end{figure}
%%%%%%%%
%

\begin{figure}[t]
	\centering
	\includegraphics[width=0.495\textwidth]{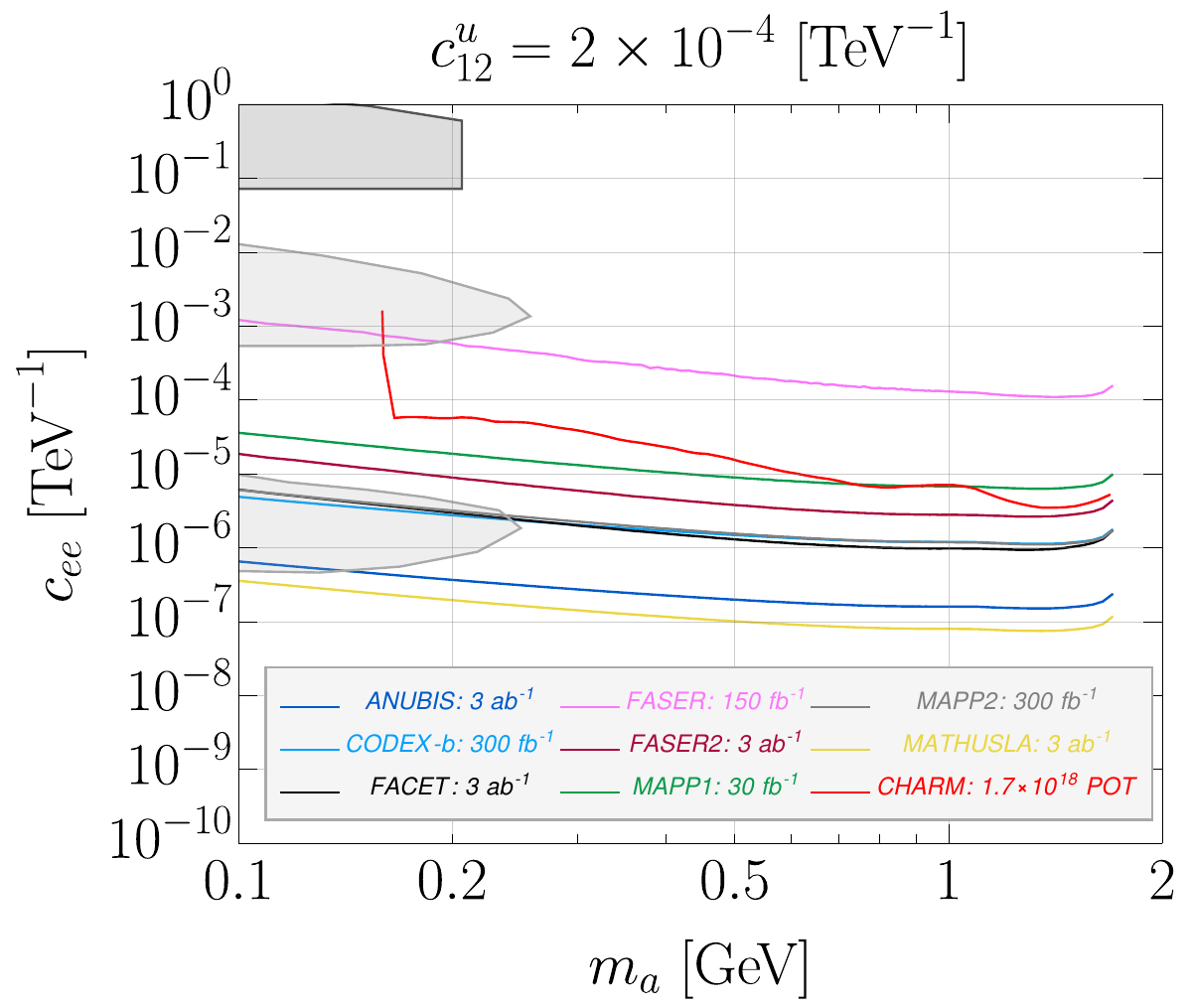}
	\includegraphics[width=0.495\textwidth]{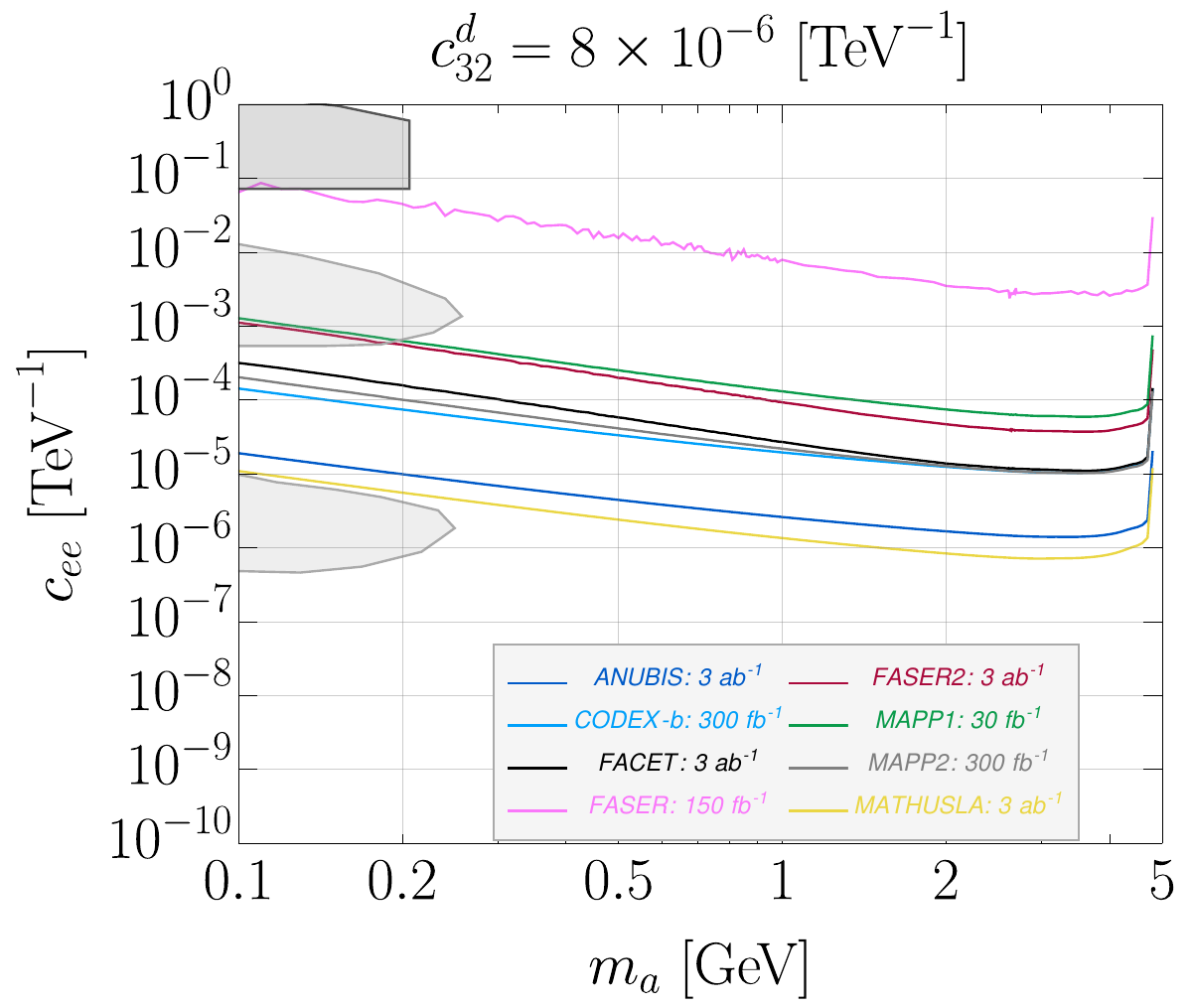}
	\caption{Reinterpretation results for ALPs produced from meson decays and decaying into a pair of electrons.
		The gray area represents the existing bounds on the coupling $c_{ee}$, obtained at E137~\cite{Essig:2010gu,Bjorken:1988as} (dark gray) and derived from supernovae~\cite{Ferreira:2022xlw} (light gray).
		Since our method does not work in the prompt regime, the corresponding bounds on higher values of $c_{ee}$ are not shown.
	}
	\label{fig:ALP}
\end{figure}

Further, to obtain the current upper bounds on $c^u_{21}$ and $c^d_{32}$, we consider the strongest existing experimental limits on the decay branching ratios of $D^0\to\pi^0 \nu\bar{\nu}$ and $B\to K\nu\bar{\nu}$, while currently there is no measurement of the decay branching ratios of $D^\pm\to \pi^\pm \nu \bar{\nu}$ and $D_s\to K \nu\bar{\nu}$.
BESIII~\cite{BESIII:2021slf} reports an upper bound of $2.1\times 10^{-4}$ on BR$(D^0\to\pi^0 \nu\bar{\nu})$ at 90\%~C.L., and Belle~\cite{Belle:2017oht} gives upper bounds of $1.9\times 10^{-5}$ on BR$(B^+\to K^+\nu\bar{\nu})$ and $1.3\times 10^{-5}$ on BR$(B^0\to K^0_S\nu\bar{\nu})$ at 90\% C.L..
In numerical studies, we multiply the latter by 2 to reach a bound of $2.6\times 10^{-5}$ on BR$(B^0\to K^0\nu\bar{\nu})$.
Plugging these numbers into Eq.~\eqref{eq:PtoPprimea1} for various possible ALP masses, we derive upper bounds on the QFV couplings $c^u_{21}$ and $c^d_{32}$.
At the end, conservatively, we choose to take $c^u_{12}=2\times 10^{-4}$ TeV$^{-1}$ and $c^d_{32}=8\times 10^{-6}$ TeV$^{-1}$ for our numerical analysis.
In Fig.~\ref{fig:BRs}, we display the branching ratios of the $D$- and $B$-meson decays discussed above, fixing the corresponding couplings to these values.
\footnote{For the $N_R$LEFT scenario with pair-$N_R$ operators discussed in Sec.~\ref{sec:pairNR}, similar plots are given in Figs.~2 and 3 of Ref.~\cite{Beltran:2022ast}, whereas for the scenario with single-$N_R$ operators considered in Sec.~\ref{sec:singleNR}, 
the reader is referred to Fig.~2 of Ref.~\cite{DeVries:2020jbs}.}
As for current bounds on the coupling $c_{ee}$, we extract supernovae limits from Ref.~\cite{Ferreira:2022xlw}, beam-dump bounds from Ref.~\cite{Essig:2010gu,Bjorken:1988as}, as well as $B$-factory bounds from BaBar~\cite{BaBar:2017tiz,Darme:2020sjf}.
The first two are shown in gray in Fig.~\ref{fig:ALP}, while the latter are not because they are too weak.

Following the reinterpretation method spelled out in Sec.~\ref{sec:method}, we show our results in Fig.~\ref{fig:ALP}, in the plane $c_{ee}$ vs.~$m_a$, fixing $c^u_{12}$ and $c^d_{32}$ 
at the above-mentioned values.
%%%%%%%%
%%%%%%%%
The left and right plots are for the benchmarks ALP-D and ALP-B, respectively.
For the charm scenario, we have studied not only the LHC far detectors, but also the CHARM search~\cite{CHARM:1985nku} which focuses on leptonic final states, while for the bottom scenario, we only take into account the far detectors, as the Belle search~\cite{Belle:2013ytx} requires a prompt lepton which is absent in the current case.
As expected, the CHARM limits are comparable to those of MAPP1.

In the upper parts of these plots, the ALPs become too short-lived to decay inside the considered detectors and therefore, the sensitivity curves should close up at the top.
However, as emphasized before, our reinterpretation method only works in the large decay length limit, and these ``prompt bounds" are hence missing.

Finally, we comment that from the plots in Fig.~\ref{fig:ALP}, it is, in principle, possible to derive bounds on the production couplings $c^u_{12}$ and $c^d_{32}$, given a different value of $c_{ee}$ than those shown in Fig.~\ref{fig:ALP}, as long as the ALP lab-frame decay length is still larger than the distance between the IP and the detector.
For instance, the left plot of Fig.~\ref{fig:ALP} shows that with $c^u_{12}=2\times 10^{-4}$~TeV$^{-1}$, MATHUSLA is sensitive to $c_{ee} \approx10^{-7}$ TeV$^{-1}$ for $m_a=1$~GeV.
For another value of $c_{ee}$, say, $10^{-6}~(10^{-8})$~TeV$^{-1}$, we easily see that this corresponds to a decay length smaller (larger) by a factor of 100, and hence a bound of $2\times 10^{-5}$ 
($2\times 10^{-3}$)~TeV$^{-1}$ on $c^u_{12}$. 
Note that these bounds will be complementary to those derived 
in Ref.~\cite{Bauer:2021mvw}.

% !TEX root = ../reinterpretation.tex
\section{Discussion and summary}\label{sec:discussionANDsummary} 

In this work, we have proposed a simple reinterpretation method for searches for long-lived particles (LLPs) produced in rare meson decays, and applied the method to reinterpret heavy neutral leptons (HNLs) in the minimal scenario mixing with electron neutrinos only, into HNLs or axion-like particles (ALPs) in different effective field theories (EFTs).

Our method allows for simple and fast reinterpretations, where no simulation, or at most simulation for only one theoretical scenario, is required, as long as the following two conditions are satisfied: i) the LLPs in different theoretical scenarios possess similar kinematics (\textit{e.g.}~all are produced from the same type of mesons or the same meson), and ii) the relevant parameter regions correspond to large decay length regime compared to the distance between the detector and the interaction point (IP).
Other factors could also affect the results, such as the spin of the LLP and the number of decay products associated with the LLP production, but they have only minor effects and are hence neglected.
Thus, with only knowledge of the production and decay rates of the LLPs, one can easily perform the reinterpretation.

In this work, for the illustrative purpose of demonstrating the reinterpretation method, we have chosen the minimal HNLs produced from charm and bottom meson decays separately, as the base model, from which we derive bounds on EFT HNLs and ALPs, which can also be produced from these mesons' decays.
For pair-$N_R$ operators, the HNLs are produced in pair via these EFT operators and decay via mixing with the electron neutrino.
For single-$N_R$ operators, we consider HNLs produced and decaying via two EFT operators of the same spinor structure but different quark flavor indices, assuming vanishing mixing with the active neutrinos.
Finally, for the ALPs, we study two benchmark scenarios, assuming a non-vanishing quark-flavor-violating coupling in each scenario leading to the ALP production, and a simple ALP coupling to a pair of electrons giving rise to the decay $a\to e^+ e^-$ at tree level.
We have focused on a series of proposed LHC far detectors such as FASER and MATHUSLA, and in addition recast two existing searches at CHARM and Belle.
For the HNLs in the EFTs, we compare our reinterpreted bounds with those published in the literature, and find generally excellent agreement.
Very few discrepancies arise, when the limits correspond to decay lengths that are not large enough, or when we relate the production and decay couplings; the reinterpretation method would be slightly off from the full-simulated results in these cases.

While in general the reinterpretation method shows excellent performance, it has its drawbacks.
Firstly, it is valid only in the large decay length limit and breaks down if the LLPs decay promptly, and hence the prompt-regime bounds for the LLPs cannot be obtained this way.
Further, since no full simulation is performed, minor effects such as the LLP's spin are missing, which could alter the bounds slightly.

We have studied both long-lived fermions (HNLs) and (pseudo-)scalars (ALPs), showing that the method is not restricted (severely) by the LLP spins.
In fact, the method is clearly also not limited to the LLPs produced from charm or bottom mesons either, and can be extended to LLPs produced in decays of pions and kaons, for example.
However, one should note that when one uses our reinterpretation method on LLPs produced from pseudo-scalar kaons, one should separate $K^0_S, K^0_L$, and $K^\pm$; this is because even though they are of similar masses resulting in similar kinematics of the LLPs produced from their decays, they have different production rates in general and their lifetimes differ by orders of magnitude.
The same note also holds for $\pi^0$ and $\pi^\pm$, for the same reasons.\footnote{For $D$- and $B$-mesons, this is not an issue, as all the mesons considered in this work can be regarded as promptly decaying compared to the scale of the distance between the detector and the IP.}
Moreover, although in this work we confine ourselves to LLPs coupled with the electron and electron neutrinos, the method should apply to LLPs coupled with the second or third generation leptons, or those coupled purely hadronically; one just needs to ensure that the corresponding final states are or can be searched for in the considered experiments.

Finally, we should mention that further examples of LLPs from meson decays include the lightest neutralinos in the R-parity-violating supersymmetry~\cite{Dedes:2001zia,Dreiner:2002xg,Dreiner:2009er,deVries:2015mfw}.
Furthermore, it should be worthwhile to explore our method for LLPs produced in direct collisions or decays of heavier particles such as the $W$-boson and the top quark.

%\bigskip
\section*{Acknowledgements}

%\medskip

We thank Juan Carlos Helo and Abi Soffer for useful discussions. G.C. would like to thank the AHEP Group at Instituto de F\'isica Corpuscular for hospitality offered while working on this project.
This work is supported by the Spanish grants PID2020-113775GB-I00
(AEI/10.13039/ 501100011033) and CIPROM/2021/054 (Generalitat
Valenciana). R.B. acknowledges financial support from the Generalitat Valenciana (grant ACIF/2021/052).
G.C. acknowledges support from ANID FONDECYT grant No. 11220237 and ANID -- Millennium Science Initiative
Program ICN2019\_044.

\appendix

% Bibliography
\bibliographystyle{JHEP}
\bibliography{reinterpretation}

\end{document}